\documentclass[aps,prd,superscriptaddress,11pt,showpacs,notitlepage]{revtex4}
	\usepackage{graphicx}
	\usepackage{amsmath,amssymb,mathrsfs}
\usepackage{epsf}
\usepackage{epsfig}
\usepackage{subfig}

\usepackage{amsmath}

\usepackage{amssymb}

\numberwithin{equation}{section}

\newcommand{\be}{\begin{equation}}
\newcommand{\ee}{\end{equation}}
\newcommand{\bea}{\begin{eqnarray}}
\newcommand{\eea}{\end{eqnarray}}

\newcommand{\bb}{\bibitem}
 
\newcommand{\eqn}{\begin{eqnarray}}
\newcommand{\eqnx}{\end{eqnarray}}

\begin{document}
\title{Neutron stars in the Bogomol'nyi-Prasad-Sommerfield  Skyrme model: mean-field limit vs. full field theory}
\author{C. Adam}
\affiliation{Departamento de F\'isica de Part\'iculas, Universidad de Santiago de Compostela and Instituto Galego de F\'isica de Altas Enerxias (IGFAE) E-15782 Santiago de Compostela, Spain}
\author{C. Naya}
\affiliation{Departamento de F\'isica de Part\'iculas, Universidad de Santiago de Compostela and Instituto Galego de F\'isica de Altas Enerxias (IGFAE) E-15782 Santiago de Compostela, Spain}
\author{J. Sanchez-Guillen}
\affiliation{Departamento de F\'isica de Part\'iculas, Universidad de Santiago de Compostela and Instituto Galego de F\'isica de Altas Enerxias (IGFAE) E-15782 Santiago de Compostela, Spain}
\author{R. Vazquez}
\affiliation{Departamento de F\'isica de Part\'iculas, Universidad de Santiago de Compostela and Instituto Galego de F\'isica de Altas Enerxias (IGFAE) E-15782 Santiago de Compostela, Spain}
\author{A. Wereszczynski}
\affiliation{Institute of Physics,  Jagiellonian University,
Lojasiewicza 11, Krak\'{o}w, Poland}

\begin{abstract}
Using a solitonic model of nuclear matter, the BPS Skyrme model, we compare neutron stars obtained in the full field theory, where  gravitational back reaction is completely taken into account, with calculations in a mean-field approximation using the Tolman-Oppenheimer-Volkoff approach. In the latter case, a mean-field-theory equation of state is derived from the original BPS field theory. 

We show that in the full field theory, where the energy density is non-constant even at equilibrium, there is no universal and coordinate independent equation of state of nuclear matter, in contrast to the mean-field approximation. We also study how neutron star properties are modified by going beyond mean field theory, and find that the differences between mean field theory and exact results can be considerable.
Further, we compare both exact and mean-field results with some theoretical and phenomenological constraints on neutron star properties, demonstrating thus the relevance of our model even in its most simple version.  

\end{abstract}
\maketitle 
\section{Introduction}
Neutron stars are macroscopic objects formed by gravitating nuclear matter. Unfortunately, it is not possible to study neutron stars within the fundamental quantum theory of strong interactions coupled to  gravity. The obvious reason is that neutron stars are made of neutrons (and a fraction of protons, electrons and perhaps some other particles) which are low energy non-perturbative excitations of QCD. Presently, it is not known how to describe non-perturbative phenomena of strongly interacting matter directly from QCD. To overcome this difficulty and study the low energy sector of QCD, one has to deal with effective field theories (EFTs) which are usually proposed (or motivated by some general arguments) rather than derived from the underlying fundamental theory. Even if one assumes an effective low energy action of QCD (for example quantum hadron dynamics, or the Nambu-Jona-Lasino model or some bag models), however, it is not possible to find neutron stars as solutions of that field theory coupled to  Einstein gravity. This most obvious way to describe neutron stars is usually too complicated. A widely accepted solution of this problem is provided by the Tolman-Oppenheimer-Volkoff (TOV) approach \cite{OV}, \cite{Tol} where the Einstein equations are solved for a perfect fluid energy-momentum tensor. To complete the set of equations, 
one usually assumes that the perfect fluid is of a barotropic nature with a barotropic equation of state (EoS),
i.e., an algebraic relation between the energy density $\rho$ and the pressure $p$. This is the place where a particular EFT finally enters. 
To derive the equation of state from a given effective action (without gravity), however, is again a non-trivial task, since EFTs at the fundamental level (i.e., using their field degrees of freedom and the corresponding  action) usually are not perfect fluids. Therefore, one has to go from a microscopic (field theoretic) description to a macroscopic description and a thermodynamic limit must be performed, which typically implies a mean-field approximation. In this manner, an average (spatially constant) value of the energy density is calculated and all deviations from spacial constancy are ignored. However, as gravity directly couples to derivative terms, any deviation from a constant value may influence some properties of neutron stars, like the maximal mass, maximal radius, or the mass-radius relation. Hence, studying this phenomenon is an important  but, at the same time, complicated issue. The main difficulty comes from the fact that a full field theoretic calculation is required if we want to understand the error introduced by a mean field approximation. So we need a low energy effective field theory of QCD which is capable of describing nuclear matter and which has a perfect fluid energy-momentum tensor already at the microscopic (field theoretic) level. These two requirements are very strong and, in principle, one may even doubt whether such a model exists, at all.

Quite surprisingly, it has been shown recently that there exists an effective low energy action, the so-called {\em BPS Skyrme model}, for which the corresponding energy-momentum tensor possesses the {\it perfect fluid} form and, therefore, no mean-field approximation has to be made. Furthermore, one can solve this model even after its coupling to gravity and find solutions representing neutron stars. Hence, this theory can also serve as a laboratory for studying the validity and accuracy of the mean field approximation, which is one of the objectives of the present work. 

Among EFTs for low energy QCD, Skyrme type models play a very prominent role \cite{skyrme}, \cite{sk}. It has been conjectured by Skyrme that baryons and nuclei may be described as a sort of {\it vorticity} in a mesonic fluid or, in a more precise mathematical language, as solitons in a mesonic effective field theory \cite{skyrme}. 
This proposal received some further motivation from the large $N_c$ expansion. Indeed, QCD in the limit of an infinite number of colours becomes equivalent to a a weakly interacting theory with only mesonic degrees of freedom \cite{thooft}. Unfortunately, the derivation of the effective mesonic theory from QCD is still an open problem and its correct form is unknown. In the simplest form, as proposed by Skyrme, the Skyrme model lagrangian reads
\be
\mathcal{L}=\mathcal{L}_2+\mathcal{L}_4+\mathcal{L}_0
\ee
where the first two terms represent the sigma model (kinetic) part and the so-called Skyrme part
\be
\mathcal{L}_2= \lambda_2 \mbox{Tr} \; \partial_\mu U \partial U^\dagger, \;\; \mathcal{L}_4=\lambda_4 \mbox{Tr} ([L_\mu, L_\nu])^2 ,
\ee
and $\mathcal{L}_0=-\lambda_0 \mathcal{U} (\mbox{Tr} \; U)$ is a non-derivative term, i.e., a potential. Usually one chooses $\mathcal{U}=\mathcal{U}_\pi = 1-\mbox{Tr} \, U$, which provides a mass for the perturbative excitations (the pions). Here, all $\lambda_n$ are non-negative, dimensionful coupling constants. Moreover, $U$ is an $SU(2)$ valued matrix Skyrme field and $L_\mu = U^\dagger \partial_\mu U$ is the left-invariant Maurer-Cartan current. The baryon number is identified as a topological charge $B$ corresponding to the conserved topological current $\mathcal{B}^\mu$
\be
\mathcal{B}^\mu = \frac{1}{24\pi^2} \epsilon^{\mu \nu \rho \sigma} \mbox{Tr} \; L_\nu L_\rho L_\sigma, \;\;\; B= \int d^3 x \mathcal{B}^0 .
\ee
From a phenomenological perspective, the original Skyrme action has been quite successful in many respects. After a semiclassical rigid body quantization (and carefully taking into account the Finkelstein-Rubinstein constrains \cite{FR}, \cite{FR-Kr}), it provides a good description of nucleons (charge $B=1$ sector) \cite{anw} as well as the deuteron \cite{braaten} and some additional light nuclei \cite{carson}. Furthermore, it gives rotational excitation bands of some light nuclei with a good qualitative as well as quantitative agreement with experimental results \cite{light} (for a possible improvement related to deformations of spinning skyrmions see \cite{rot}). As a very recent example, the model allows to model the structure and properties (e.g., the so-called Hoyle state) of Carbon 12 \cite{carbon}. 

This notable success of the original Skyrme model applied to excited states of the lowest nuclei has to be contrasted with its inability to describe the proper binding energies of atomic nuclei. Skyrmions with higher topological charges have binding energies approximately 10 times bigger than experimentally measured. Another important issue is the fact that skyrmions with large baryon number form crystals, which again is in contrast to the liquid behaviour of nuclear matter. In order to cure the binding energy problem, two possible generalizations of the original Lagrangian to a so-called {\it near BPS} Skyrme model have been proposed. Although they explore the same idea - to bring the model close to a BPS theory - they achieve it in  rather different ways. One possibility is to flow the usual (static) Skyrme model into the conformal Yang-Mills $SU(2)$ theory in four dimensional Euclidean space by the inclusion of (infinitely many) higher vector mesons \cite{SutBPS} (see also \cite{rho}). Alternatively, one can add a dominating BPS submodel based entirely on the mesonic Skyrme field \cite{BPS}
\be
\mathcal{L}_{BPS} \equiv \mathcal{L}_{06}=\mathcal{L}_6+\tilde{\mathcal{L}}_0,
\ee
where the derivative dependent sextic term is just the baryon current squared 
\be
\mathcal{L}_6= -(24\pi^2)^2 \lambda_6 \mathcal{B}_\mu \mathcal{B}^\mu ,
\ee
and $\tilde{\mathcal{L}}_0$ is again a potential.
Then, the full {\it near BPS Skyrme} model reads
\be \label{near-BPS}
\mathcal{L}=\mathcal{L}_6+\tilde{\mathcal{L}}_0+ \epsilon( \mathcal{L}_2+\mathcal{L}_4+\mathcal{L}_0 )
\ee
where $\epsilon$ is a small parameter (see also \cite{Marl}, \cite{Sp2}; for the influence of the sextic term on the usual
Skyrme model, see, e.g., \cite{sextic}). 
It has been shown that this improved Skyrme action, already in the exact BPS limit $\epsilon \to 0$, leads to very accurate binding energies if one takes into account the semiclassical quantization of (iso)-rotational degrees of freedoms, the Coulomb interaction as well as the isospin symmetry breaking \cite{nearBPS}. Furthermore, the BPS Skyrme static energy functional is invariant under volume preserving diffeomorphisms (VPDs) on physical space, which are the symmetries of a perfect fluid. In addition, the energy-momentum tensor of the BPS Skyrme model is the energy-momentum tensor of a perfect fluid. This solves the second, before mentioned, long standing problem in the Skyrme framework. All these results support the conjecture that the near BPS Skyrme model might be the right low-energy EFT for the description of nuclear matter. One has to underline that already the BPS part of the full near BPS model gives a very good approximation to some static properties of nuclear matter (masses, binding energies). This is important since the BPS sector is {\it solvable} which renders computations possible. We remark that another variant of the Skyrme model leading to low binding energies, where a 'repulsive' potential is added to the standard Skyrme action, and whose existence is, again, based on topological energy bounds \cite{harland}, \cite{top-bounds}, has been proposed and investigated recently in \cite{har-gil-sp}.  

The solvability of the BPS Skyrme model allows to find solitonic solutions after coupling to gravity \cite{star}. The important point is that the obtained gravitating skyrmions (neutron stars) are solutions of the field equations with the Einstein gravity fully taken into account. This means that no assumption on an EoS was made. Instead, the Skyrme matter acts as a source (energy-momentum tensor) for the Einstein equations (for previous applications of skyrmions in the context of neutron stars and as a source for GR equations see \cite{bizon} - \cite{gravSk}). The validity of this approach follows from the observation that, as mentioned above, the energy-momentum tensor is of the perfect fluid type and no thermodynamical limit has to be taken to reach a fluid regime. An (algebraic) on-shell equation of state may be computed after solving the equations, but it turns out that this on-shell EoS is not unique but depends on the total mass of the neutron star, i.e., on the specific solution. This is related to the fact that the original EoS in the BPS Skyrme model (without gravity) is not of an algebraic form $\rho = \rho (P)$ (i.e., the perfect fluid is not barotropic), but depends on spatial coordinates i.e., $\rho=\rho(r,P)$ since the energy density $\rho$ is typically not spatially constant. 
$P$, on the other hand, is an integration constant in the field theory context, so it is constant by definition.

The objective of the present paper is twofold. We shall continue the investigation of neutron star properties within the BPS Skyrme model commenced in \cite{star}, and we want to compare the full field theory and gravity computations with the mean-field TOV approach. To accomplished this second aim, we have to derive an algebraic (global) EoS from the correct (coordinate dependent) EoS by a sort of mean-field approximation. 

The outline of the paper is the following. In the next section, we briefly summarize some thermodynamical properties of the BPS Skyrme model. In section III it is shown how an average, mean-field equation of state (MF-EoS) may be derived from the exact EoS. In section IV we describe the results of our numerical neutron star calculations and compare the neutron star properties obtained in the full field theoretical and gravity set-up with the TOV approach, where the MF-EoS is used. In section V, we discuss our results, with emphasis on the mass-radius relation ($M(R)$ curve), and including the TOV inversion issue. We leave the comparison with other approaches to section VI where, in particular, we compare both to generic and to Skyrme-model related  results. Finally, in section VII we summarize our conclusion and give some outlook. 
\section{Equation of state in the BPS Skyrme model}
One of the most striking features of the BPS Skyrme model is the fact that, at the same time, it provides both a description in terms of microscopic degrees of freedom (mesonic fields in the BPS Skyrme action) and a description by macroscopic quantities, i.e., thermodynamical functions \cite{term}. Both descriptions agree completely in the sense that the field theoretical energy $E$ and pressure $P$ together with the geometrical volume $V$ obey the standard thermodynamical relation 
\be
P=-\frac{d E}{d V}.
\ee 
The main consequence of this result is that, in the BPS Skyrme model, we are all the time in a thermodynamical limit. In other words, no additional thermodynamical limit (for example a kind of mean field approximation) has to be performed to reach a macroscopic description in the language of the proper thermodynamical functions. This is a profound difference between the BPS Skyrme model and any other low energy effective theory of Quantum Chromodynamics.
Below we briefly summarize some thermodynamical properties of the BPS Skyrme model \cite{term}.
\\
First of all, the energy-momentum tensor of the BPS Skyrme model has the form of the energy-momentum tensor of a perfect fluid (for the moment we consider the case of flat Minkowski space, i.e., without gravity)
\be
T^{\mu \nu} = (p+\rho) u^\mu u^\nu - p \eta^{\mu \nu}
\ee
where the energy density and pressure for static configurations, which are the ones relevant for us (and where $u^\mu = (1,0,0,0)$), read
(for convenience, we introduce the new coupling constants $\lambda^2 = (24)^2 \lambda_6$ and $\mu^2 = \lambda_0$)
\be
\rho = \lambda^2 \pi^4 \mathcal{B}_0^2+\mu^2\mathcal{U}, \;\;\; p=\lambda^2 \pi^4 \mathcal{B}_0^2-\mu^2\mathcal{U} .
\ee
From the conservation equation $\partial_\mu T^{\mu \nu}=0$ we get $p=P=const.$ Hence, the pressure inside BPS skyrmions (skyrmionic matter) is constant. 
\\
Moreover, the equation which defines the pressure is, in fact, a first integral of the full second order equation of motion. It provides solutions (field configurations with nontrivial baryonic charge) with non-zero (external) pressure. In particular, for the axially symmetric ansatz 
\be
U=\cos \xi + i \sin \xi \vec{n} \cdot \vec{\tau}
\ee  
where $\vec{\tau}$ are Pauli matrices and 
\be
\xi =\xi (r), \;\;\; \vec{n}=(\sin \theta \cos B \phi, \sin \theta \sin B \phi, \cos \theta)
\ee
we get ($\xi_r \equiv \partial_r \xi$)
\be \label{BPS-Pnot0}
\frac{|B|\lambda}{2r^2} \sin^2 \xi \xi_r= - \mu \sqrt{\mathcal{U}+\frac{P}{\mu^2}} 
\ee
(the usual BPS equation is obtained as a zero pressure condition
\cite{bazeia}).
Solving this equation with topologically nontrivial boundary conditions ($\xi(0)=\pi, \; \xi(R)=0$), we get a profile of the skyrmion. Here $R$ is the geometric radius of the soliton. Inserting it back into the energy density we get a spatially dependent function which also changes if the pressure is varied, i.e.,
\be
\rho = \rho (r, P) .
\ee
This formula can be regarded as an on-shell density-pressure equation of state in the BPS Skyrme model for spherically symmetric nuclear matter. 
\\
Another important quantity considered in the context of nuclear matter is the particle density. Here, it is equivalent to the baryon topological charge density $\mathcal{B}_0$. The corresponding equation of state may be easily read-off from the components of the energy-momentum tensor. Namely, 
\be
 \mathcal{B}_0 (r,P)= \frac{1}{\lambda \pi^2} \sqrt{\rho(r,P)+P}
\ee
where the plus sign has been assumed as we have baryons (not anti-baryons). This, again, is the particle number - pressure equation of state which together with the above energy density - pressure EoS completely defines the system. Obviously, in our construction both equations of state
follow from the BPS Skyrme action. 
\\
Furthermore, one can find the total energy and the geometrical volume 
\begin{equation}
E(P)=2\pi \lambda \mu |B|\tilde{E}, \;\;\; V=2\pi |B| \frac{\lambda}{\mu} \tilde{V}
\end{equation}
where
\begin{equation} \label{tilde-E}
\tilde{E}=\int_0^\pi d\xi \sin^2 \xi \frac{2\mathcal{U}+\tilde{P}}{\sqrt{\mathcal{U}+\tilde{P}}}, \;\;\; \tilde{V}=\int_0^\pi d\xi \sin^2 \xi \frac{1}{\sqrt{\mathcal{U}+\tilde{P}}} .
\end{equation}
Here, the base space integrals defining $E$ and $V$ in the usual way may be converted into the (solution-independent) target space integrals (\ref{tilde-E}) with the help of the first-order (BPS) equation (\ref{BPS-Pnot0}). 
Further,
 $\tilde{P}=P/\mu^2$. In contrast to the density, the total energy and the volume are related to the pressure in a global (coordinate independent) manner. So, some well-defined energy-pressure $E=E(P)$ as well as volume-pressure $V=V(P)$ equations of state do exist. It should be noted that the energy and volume can be found without knowing a particular solution. They are given by integrals of some functions of the potential over the target space.

\section{Mean-field equation of state - a hadronic bag model}
As underlined already, it is an inherent property of the BPS Skyrme model (and undoubtedly its near BPS extensions) that the energy density of skyrmionic matter is not constant - except for the step function potential. On the other hand, the model allows to describe solutions with non-zero external pressure, and the pressure is always {\it constant} inside skyrmions (nuclear matter). Therefore, the energy density is a space (radial) dependent function which changes if a non-zero pressure is imposed,  
\be
\rho=\rho(r, P), \;\; P=const.
\ee
It should be emphasized, again, that this is the proper equation of state of skyrmionic matter in the BPS Skyrme model. Hence, in general, the pressure and the energy density are not related by an algebraic equation (which does not depend on coordinates) i.e., there is no (barotropic) equation of state of the form 
\be
\rho = \rho (P).
\ee
This may be a source of some difficulties if one wants to compare the thermodynamical properties of the BPS Skyrme model with the usually used density-pressure equations of state derived in other effective field theories. 
However, it is possible to cast the equation of state obtained in the BPS Skyrme model into such an algebraic (global) form. This requires a simple averaging procedure, which may be interpreted as a mean field approximation. The obvious definition of the average energy density is
\begin{equation}
\bar{\rho}(P)=\frac{E(P)}{V(P)}
\end{equation}
and the resulting mean field equation of state (MF-EoS) is just a function of some target space integrals (averages) 
\begin{equation}
\bar{\rho}(P)=\mu^2 \frac{\left<\frac{2\mathcal{U}}{\sqrt{\mathcal{U}+P/\mu^2}}\right>}{\left<  \frac{1}{\sqrt{\mathcal{U}+P/\mu^2}} \right>} +P.
\end{equation}
Here 
\be
\left< F(\mathcal{U})\right>  \equiv \frac{1}{{\rm Vol}_{{\mathcal S}^3}} \int {\rm vol}_{{\mathcal S}^3} F(\mathcal{U}) = \frac{2}{\pi} \int_0^\pi d\xi \sin^2 \xi F(\mathcal{U})
\ee
where we use that, as a manifold, the target space SU(2) may be identified with the unit three-sphere ${\mathcal S}^3$. Further, 
${\rm Vol}_{\mathcal{S}^3} = 2\pi^2 $ is the volume of the (target space) unit three-sphere, and ${\rm vol}_{\mathcal{S}^3}$ the corresponding volume form. The right hand side follows because the potential ${\mathcal U}$, by assumption, depends on the Skyrme field only via the field variable $\xi$. 

For later convenience, we also want to define the average baryon density as baryon number divided by volume,  i.e., 
\be \label{av-b-dens}
\bar n_{B} = \frac{B}{V} = \frac{\mu}{\pi^2 \lambda \left<  \frac{1}{\sqrt{\mathcal{U}+P/\mu^2}} \right>}.
\ee

Notice that it is not necessary to know explicit soliton solutions of the model to find the MF-EoS. Instead, it is encoded in geometrical (target space) average values.
Obviously, a particular form of the MF-EoS follows from a particular potential. However, some general observations can be easily made. 
\subsection{High pressure limit - a hadronic bag model.}
For large values of the pressure we find the following expansion (up to the first two terms; we use $\left< 1\right> = 1$)
\be
\bar{\rho}(P)=\mu^2 \frac{\left< 2\mathcal{U}\right> }{\left< 1\right> }+P= P+ 2 \mu^2  \left< \mathcal{U}\right> .
\ee
This is an EoS of a bag type matter with the asymptotical bag constant $B_\infty$
\be
\bar{\rho}(P)=P+B_\infty, \;\;\; B_\infty=2\mu^2 \left< \mathcal{U}\right> 
\ee
(in the neutron star context, this EoS is known as the "maximally compact EoS"; see below). 
Hence, at least for high pressure and from the average MF-EoS point of view, matter described by the BPS Skyrme model behaves as a matter in a bag type model, where the asymptotic bag constant is given again by a target space average of the potential. In comparison, the MIT bag model leads to the following linear equation of state \cite{MITbag}
\be
P=\frac{1}{3} (\bar{\rho}-4B_{\rm MIT}) .
\ee
 As the Skyrme model is based entirely on a mesonic matrix field with all quark contributions assumed to be integrated out, we seem to have a kind of {\it hadronic bag model} instead of the typical quark bag model - the BPS Skyrme model describes nuclear matter, not quark matter. Moreover,  in the MIT bag model the proportionality constant is $1/3$ instead of 1. 
 
 Nevertheless, in the high density (high pressure) limit, the EoS of the BPS Skyrme model coincides with the high-density EoS of several other important models of nuclear matter. First of all, let us observe that in the limit of very high pressure, even the exact (non-constant) energy and baryon number densities become approximately constant ($\rho \sim \bar \rho$ and  ${\cal B}_0 \sim \bar n_B \equiv (B/V)$) and lead to the approximate EoS
 \be
 \bar \rho = P = \pi^4 \lambda^2 \bar n_B^2.
 \ee
 But this is exactly the high-density limit of the EoS of the Walecka model \cite{walecka} (for a good review see, e.g.,  \cite{schmitt}), where the constant $\pi^4 \lambda^2$ is replaced by $(1/2)(g_\omega^2/m_\omega^2)$ (here $g_\omega$ and $m_\omega$ are the coupling constant and mass of the vector meson of the Walecka model). 
 
 Secondly, the same high-density EoS $\bar\rho \sim P$ also occurs in a modification of the MIT bag EoS, where interactions with higher mesons are taken into account. Specifically, for high $P$, contributions from the vector meson interactions always win over the usual free fermion gas part. Therefore, asymptotically one gets again $P \sim \bar{\rho}$ \cite{thomas}, coinciding with the result obtained in the BPS Skyrme model. An explanation of this agreement may be related to the fact that the interaction used in \cite{thomas} in the derivation of the EoS is effectively described by a Nambu-Jona-Lasinio (NJL) interaction term, i.e., the square of the quark vector current 
 \be
 J^\mu_{V} = \bar{q} \gamma^\mu  q.
 \ee
 Further, the quark vector current is, up to a constant, just the baryon density current, and the NJL interaction term is, therefore, equivalent to the sextic (baryon density current squared) term in the BPS Skyrme model, although the field parametrization of the current is quite different in the two models - a Noether current based on quark fields in the NJL case, and a topological current based on the Skyrme field in the BPS Skyrme model. As the vector channel gives the leading contribution for the MIT bag EoS in the large pressure limit, the effective action is also dominated by this term. Thus, both models coincide in the high pressure regime, even though the baryonic current is defined by very different degrees of freedom: quark and mesonic, respectively. This observation may point towards the interesting possibility that the BPS Skyrme model is a solitonic realization of a vector MIT bag model, as conjectured by M. Nowak and M. Rho.
\subsection{Small pressure limit}
For small pressure $P$, the power series expansion in $P$ does not work. The reason is that one gets target space averages of arbitrary negative powers of the potential, $\left<  \mathcal{U}^{-a}\right>$, where $a>0$. As the potential has at least one zero (the vacuum) we get singularities for sufficiently large $a$, and the expansion breaks down. 
\\
In general the small $P$ behaviour is 
\be
\bar{\rho}(P)=2\mu^2 \frac{<\mathcal{U}^{1/2}>}{<\mathcal{U}^{-1/2}>}+ f(P)
\ee
where $f(P)$ is a non-polynomial function of $P$, such that $f(P)=0$ as $P \rightarrow 0$. Its particular form strongly depends on the potential. 
We may interpret the first constant as an equilibrium bag constant (i.e., the average energy density of nuclear matter at equilibrium or nuclear saturation density)
\be
B_0=2\mu^2 \frac{<\mathcal{U}^{1/2}>}{<\mathcal{U}^{-1/2}>}.
\ee
Let us notice that this expression is nonzero only if $<\mathcal{U}^{-1/2}>$ is not singular. This condition is equivalent to the fact that the geometrical volume $V$ (which, as we already know, is also the thermodynamical volume) is finite at $P=0$ , which as a consequence leads to the requirement that all skyrmions have to be compactons. However, compact solitons are obtained only if the approach to the vacuum of the potential is weaker than $\xi^6$, i.e., $\lim_{\xi \to 0}{\cal U} \sim \xi^\alpha$ for $\alpha <6$. In this paper, we shall consider only potentials which satisfy this condition.


{
\subsection{Causality}
There is one physically important condition which our MF-EoS should obey. Namely, the speed of sound 
\be
\frac{1}{v^2} = \frac{\partial \bar{\rho}}{\partial P} 
\ee
has to be smaller than or equal to the speed of light as required by causality, 
\be
v \leq 1.
\ee
Therefore,
\be
\frac{\partial \bar{\rho}}{\partial P} \geq 1
\ee
which leads to the following requirement for the target space averages
\be
\left<\mathcal{U} \left(\mathcal{U} +\frac{P}{\mu^2} \right)^{-1/2}\right>
\left<\left(\mathcal{U} +\frac{P}{\mu^2} \right)^{-3/2}\right> -\left<\mathcal{U} \left(\mathcal{U} +\frac{P}{\mu^2} \right)^{-3/2}\right> 
\left<\left(\mathcal{U} +\frac{P}{\mu^2} \right)^{-1/2}\right>\geq 0
\ee
This formula provides admissible potentials. 
We will check this condition for all potentials considered below. 
\\
Furthermore, using the Chebyshev integral inequality it is possible to prove that the asymptotic bag constant $B_\infty$ is always greater than or equal to the equilibrium bag constant $B_0$,
\be
B_\infty \geq B_0.
\ee
\subsection{Examples}
\subsubsection{The step function potential}
Let us now consider three particular examples. First of all, we analyze the step function potential 
\be
\mathcal{U}=\Theta ({\rm Tr}(1- \, U)).
\ee
From a phenomenological point of view, this potential is not the proper choice. It leads to an unphysically large compression modulus \cite{term}. Nevertheless, it is still interesting to consider this case, as for the step function potential the exact equation of state precisely agrees with its mean-field average version. In more physical terms, this implies that the step function potential corresponds to the extreme case leading to completely flat (constant) energy and baryon number densities.  Indeed, both exact and MF approaches give the following EoS
\be
\rho = P+2\mu^2.
\ee
Hence $B_\infty = B_0=2\mu^2$ and $\bar{\rho}=\rho$. The reason why these two approaches lead to the same answer is the fact that the energy density is a constant function. Thus, the mean field (average) energy density is equal to the energy density computed from the BPS equation. It is also clear why, asymptotically for large pressure, the average MF-EoS for any admissible potential looks the same as for the step function potential. From the BPS equation 
\be
\frac{|B|\lambda}{2r^2} \sin^2 \xi \xi_r = -\mu\sqrt{\mathcal{U}+\frac{P}{\mu^2}}
\ee
we see that for $P >> \mu^2$, the right hand side is effectively equal to $-\mu \sqrt{P}$. This follows from the observation that the original potential takes values between the vacuum value $\mathcal{U}(\xi=0)=0$ and a maximum value $\mathcal{U}(\xi =\pi)$, which is negligible for sufficiently large pressure. Hence, in this limit, the nontrivial and field dependent right hand side effectively behaves as  a field independent (constant) quantity. Therefore, also the left hand side, i.e., the baryon charge density as well as the corresponding energy density are constant, and the resulting EoS must correspond to the EoS of the step function potential.
\subsubsection{The Skyrme potential $\mathcal{U}=\mathcal{U}_\pi$}
Another obvious choice is the Skyrme potential originally used to provide masses for (pionic) field perturbations
\be
\mathcal{U}=\mathcal{U}_\pi=2 \sin^2 \frac{\xi}{2} =1-\cos \xi \equiv 2h
\ee
(where we defined the new field variable $h$ for later convenience).
In this case, the energy density obtained from the BPS equation is not constant and the mean-field averaging leads to a different MF-EoS
\begin{equation}
\bar{\rho}=\frac{\mu^2}{5} \left( 2-3 \frac{P}{\mu^2} +\frac{6}{1+\frac{P}{\mu^2} \left( 1- \frac{K \left[ \frac{2}{2+P/\mu^2} \right]}{E \left[ \frac{2}{2+P/\mu^2} \right]}  \right)} \right)
\end{equation}
where $K$ and $E$ are the complete elliptic integrals of the first and second kind, respectively.  Here, the bag constants are
\be 
B_\infty= 2 \mu^2, \;\;\; B_0=\frac{8}{5} \mu^2.
\ee
The expansion at zero pressure gives
\be
\bar{\rho}=\frac{\mu^2}{5} \left( 8 -  \frac{P}{\mu^2}  \ln \frac{P}{2 \mu^2}  \right)
\ee
where the subleading terms have been omitted. Then, $\frac{\partial \bar{\rho}}{\partial P} =\infty$ at $P=0$. As the MF-EoS is a monotonous function of the pressure and tends to $\bar{\rho}=P+B_\infty$ (see Fig. 1a) we can conclude that the speed of sound is always smaller than 1 and the model satisfies the causality condition.  
\begin{figure}
\subfloat[]{
\label{fig:subfig:a}           
\includegraphics[height=4.5cm]{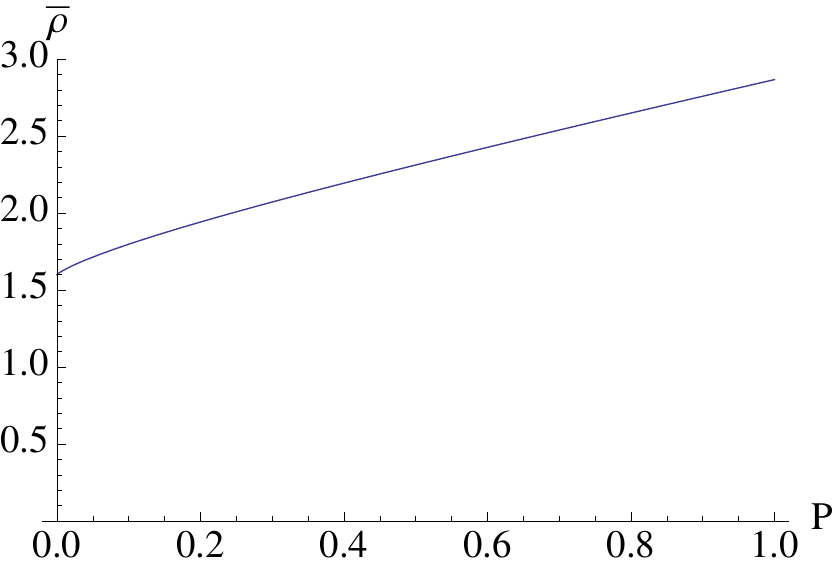}}
\hspace{0.3in}
\subfloat[]{
\label{fig:subfig:b}           
\includegraphics[height=4.5cm]{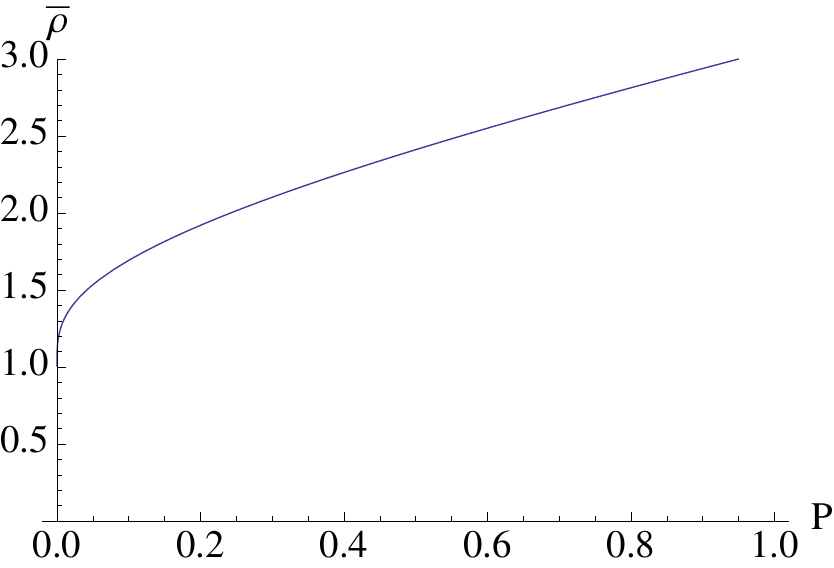}}
\caption{(Color online) The mean-field EoS for $\mathcal{U}_\pi$ and $\mathcal{U}_\pi^2$. Here $\mu^2=1$.}
\end{figure}
\subsubsection{The Skyrme potential squared $\mathcal{U}=\mathcal{U}_\pi^2$}
The last example we will explore further in the next section is the Skyrme potential squared
\be
\mathcal{U}=\mathcal{U}_\pi^2.
\ee
Then, we find the following exact expression for the MF-EoS 
\begin{equation}
\bar{\rho}=\mu^2 \left( \frac{P}{\mu^2}+ \frac{5}{2}\, \frac{{}_3F_2 [\{ \frac{1}{2}, \frac{7}{4}, \frac{9}{4} \}, \{ \frac{5}{2}, 3 \}, -\frac{4 \mu^2}{P} ]  }{{}_3F_2 [\{ \frac{1}{2}, \frac{3}{4}, \frac{5}{4} \}, \{ \frac{3}{2}, 2 \}, -\frac{4 \mu^2}{P} ] }  \right)
\end{equation}
where ${}_pF_q[\{a_1,..,a_p\}, \{b_1,..,b_q\},z] $ is a generalized hypergeometric function. Here
\be
B_\infty= \frac{5}{2} \mu^2,\;\;\;\; B_0=\mu^2.
\ee
It can be shown that the model obeys the causality requirement. In Fig. 1b we plot this MF-EoS. As is clearly visible, it tends quite rapidly to a linear density-pressure function. Hence, the EoS approaches the EoS for the step function potential rather quickly.  
\section{TOV vs. full gravitating field theory}
\subsection{Full field theoretical computations}
The BPS Skyrme model in curved space-time has the following form (for a more detailed analysis see \cite{star})
\be
S_{06}=\int d^4 x |g|^{\frac{1}{2}} \left( -\lambda^2 \pi^4 |g|^{-1} g_{\mu \nu} \mathcal{B}^\mu \mathcal{B}^\nu - \mu^2 \mathcal{U} \right) .
\ee
As in the flat space case, the corresponding energy-momentum tensor has a perfect fluid form, which for static solutions and for a diagonal metric reads
\be
T^{00}=\rho g^{00}, \;\;\; T^{ij} = - p g^{ij}
\ee
where now the pressure and energy density are metric dependent functions
\be
\rho= \lambda^2\pi^4 |g|^{-1} g_{00} \mathcal{B}^0 \mathcal{B}^0 + \mu^2 \mathcal{U}
\ee
\be
p= \lambda^2\pi^4 |g|^{-1} g_{00} \mathcal{B}^0 \mathcal{B}^0 - \mu^2 \mathcal{U}.
\ee
In order to find neutron stars in the BPS Skyrme model (in a full field theoretic calculation) one has to solve the Einstein equations 
\be
G_{\mu \nu} = \frac{\kappa^2}{2} T_{\mu \nu}
\ee
(here $G_{\mu\nu}$ is the Einstein tensor and $\kappa^2 = 16 \pi G = 6.654 \cdot 10^{-41} \, {\rm fm} \, {\rm MeV}^{-1}$) 
with the energy-momentum tensor provided by the BPS action and in a spherically symmetric metric 
\be
ds^2 = {\bf A} (r) dt^2-{\bf B} (r) dr^2 - r^2 (d\theta^2 +\sin^2 \theta d\phi^2).
\ee
The key point is that the full field theoretical equations and the Einstein equations are compatible with the metric ansatz together with the previously introduced axially symmetric ansatz for the Skyrme field. This allows to reduce the equations to a set of three ordinary differential equations: two which couple the Skyrme field (profile) $h\equiv (1/2) (1-\cos \xi)$ and ${\bf B}$, while the third one determines ${\bf A}$ in terms of $h$ and ${\bf B}$ 
\bea
\frac{1}{r} \frac{{\bf B}'}{{\bf B} } & =- & \frac{1}{r^2} ({\bf B} -1) +\frac{\kappa^2}{2} {\bf B}  \rho \label{eq1} \\
r({\bf B}  p)' &=& \frac{1}{2} (1-{\bf B} ) {\bf B}  (\rho +3p) +\frac{\kappa^2}{4} r^2 {\bf B}^2 (\rho-p)p \label{eq2}\\
\frac{{\bf A}'}{{\bf A} } &=&\frac{1}{r} ({\bf B} -1) +\frac{\kappa^2}{2} r {\bf B}  p \label{eq3}
\eea
where now
\be \label{rho-p-r}
\rho = \frac{4B^2\lambda^2}{{\bf B} r^4} h(1-h) h_r^2+\mu^2 \mathcal{U}(h), \;\;\; p=\rho-2\mu^2 \mathcal{U}(h).
\ee
These equations are solved for topologically non-trivial boundary conditions for the Skyrme profile
\be
h(r=0)=1, \;\;\; h(r=R)=0
\ee
together with natural conditions for the metric field and the pressure (which is equivalent to a condition for $h_r(r=R)$)
\be
{\bf B}(r=0)=1, \;\;\; p(r=R)=0.
\ee
A detailed discussion of these numerical computations is provided in \cite{star}. We want to emphasize that in this field theoretical approach with the gravitational interaction fully taken into account no fixed (unique) EoS has been assumed. On the contrary, the energy density $\rho$ and pressure $p$ are both metric dependent functions and, therefore, it is not expected that any {\it unique} EoS for {\it all} solution may exist. 
We remark that a similar geometry-dependent (so-called quasi-local) EoS has been used in
an analysis of gravastars and neutron stars with
anisotropic matter, see, e.g., \cite{cat} - \cite{silva}.
However, for the assumed ansatz, which is natural for neutron stars, one can derive an on-shell  EoS. Indeed, as $p$ and $\rho$ (after finding a particular solution) are functions of only one coordinate $r$, we may eliminate it and derive a relation $\rho=\rho(p)$. It should be stressed that this relation has to be found for each solution (each neutron star with a given mass) independently. One important result is that in this full field theoretical approach, there is no unique EoS (even for the spherically symmetric gravitating skyrmions). The obtained equation of state not only relates local quantities (pressure and energy density) but, in addition, depends on a global parameter i.e., the total mass $M$ (or equivalently the baryon charge) of the neutron star, i.e., 
\be
\rho=\rho (p, M)
\ee
Numerically, it has been found that the EoS is of a polytropic form $p=a\rho^b$, where the two parameters $a$ and $b$ depend on the total mass \cite{star}. 
\subsection{Mean-field approximation computations}
The same set of Einstein equations (\ref{eq1}), (\ref{eq2}) can be used for the usual mean-field TOV computation. The only difference is that now the matter field is "averaged" in the sense that instead of the field dependent energy density and pressure we deal with their mean-field versions related by the mean-field equation of state (MF-EoS) $\bar{\rho}=\bar{\rho} (p)$ introduced above. Computationally it means that we solve (\ref{eq1}), (\ref{eq2}) where $p$ and $\bar \rho$ (which replaces $\rho$) are treated as independent functions, and a fixed (spatially independent) MF-EoS $\bar{\rho}=\bar{\rho} (p)$ is assumed, which closes the system. The only initial condition is ${\bf B}(r=0)=1$.
\\
Obviously, for the step-function potential both approaches are exactly the same, as the full field theoretical EoS is equal to its mean-field version.  

\subsection{Parameter values and initial conditions}
Before performing the numerical calculations, we have to choose numerical values for the coupling constants $\lambda$ and $\mu$ of the BPS Skyrme model for the three potentials ${\cal U} = \Theta (h)$, ${\cal U} = {\cal U}_\pi = 2h$ and ${\cal U} = {\cal U}_\pi^2 = 4h^2$ we want to consider. We shall determine these values by fitting the BPS skyrmions to properties of nuclear matter. In \cite{star}  we fitted to the mass of the helium nucleus and to the nucleon radius for simplicity, but here we prefer to fit to the binding energy per nucleon of infinite nuclear matter $E_{\rm b} = 16.3 \, {\rm MeV}$ and to the  nuclear saturation density (baryon density of nuclear matter at equilibrium at zero pressure), $n_0 = 0.153 \; {\rm fm}^{-3}$ \cite{schmitt}, because infinite nuclear matter is a more appropriate choice for neutron stars. (In any case, the differences in the values of physical observables induced by the two fits are rather small and typically do not exceed 15\%). With the nucleon mass $E_{\rm n} = 939.6 \; {\rm MeV}$, the soliton energy per nucleon is $E_{B=1} = E_{\rm n} - E_{\rm b} = 923.3 \, {\rm MeV}$. Further, $V_{B=1} = (1/0.153)\,{\rm fm}^3$. Then, using the expressions for energy and volume at zero pressure (at nuclear saturation), the fit in the three cases leads to
\be \label{param-pot-theta}
\Theta (h): \; \;  E = 2\pi^2 B\lambda \mu  ,  \; \;  V = \pi^2 B\frac{\lambda}{\mu}  \; \; 
 \Rightarrow \; \;  \lambda^2 = 30.99 \; {\rm MeV} \,{\rm fm}^3  , \; \;  \mu^2 = 70.61 \; {\rm MeV}\, {\rm fm}^{-3}  
\ee
\be \label{param-pot-2}
{\cal U}_\pi: \; \;  E = \frac{64 \sqrt{2}\pi}{15} B\lambda \mu  ,  \; \;  V = \frac{8}{3}\sqrt{2}\pi B\frac{\lambda}{\mu}  \; \; 
 \Rightarrow \; \;  \lambda^2 = 26.88 \; {\rm MeV} \,{\rm fm}^3  ,  \; \;  \mu^2 = 88.26 \; {\rm MeV}\, {\rm fm}^{-3}  
\ee
\be \label{param-pot-4}
{\cal U}_\pi^2: \; \;  E = 2\pi^2 B \lambda \mu ,  \; \;  V = 2\pi^2 B \frac{\lambda}{\mu} \;  \; 
\Rightarrow \; \;  \lambda^2 = 15.493 \; {\rm MeV} \,{\rm fm}^3  , \; \;   \mu^2 = 141.22 \; {\rm MeV}\, {\rm fm}^{-3}  
 \ee
It is of some interest to consider the "initial conditions" at $r=0$ and the resulting free integration constants in the two cases (exact field theory, and TOV equations using the MF-EoS, respectively). At first sight, it seems that we have one free constant at $r=0$ in both cases, which may be chosen to be the value of $\rho$ ($\bar \rho$) at the center, i.e., $\rho (0)$ or $\bar \rho (0)$, respectively. Here, in the exact field theory case, $\rho (0)$ is related to the second Taylor coefficient $h_2$ (where $h = 1-h_2 r^2 + {\bf O}(r^3)$) via $\rho (0) = 16 B^2 \lambda^2 h_2^3 + \mu^2 {\cal U}(1)$, as may be checked easily. It seems that for each initial value of $\rho (0)$ or $\bar \rho (0)$ we just have to integrate up to a value $r=R$ where $p(R)=0$ which defines the surface of the resulting neutron star, such that there is one solution per initial value (at least as long as the initial values are not too big, such that the condition $p(R)=0$ can be satisfied for some $R$). 

It turns out, however, that the initial value $\rho (0)$ may not be chosen arbitrarily in the exact field theory case. The reason is that, at the surface $r=R$, the Skyrme field $h$ must take its vacuum value $h=0$ where the potential is zero, ${\cal U}(0)=0$. But this immediately implies that, at the surface $r=R$, the energy density must be zero, too, i.e., $\rho (R)=0$. It then follows from Eq. (\ref{eq2}) that a metric function ${\bf B}$ which is nonsingular at the surface may exist only if $p$ satisfies the condition $p'(R)=0$, as well. In other words, we have to impose the condition $p'(R)=0$ in addition to $p(R)=0$, and, in general, both conditions may be satisfied simultaneously at most for a discrete set of initial values $\rho (0)$. We may, nevertheless,  find different solutions with different neutron star masses, because in the exact field theory the baryon number $B$ enters as an additional free parameter, see Eq. (\ref{rho-p-r}). Concretely, we find one solution (i.e. one initial condition $\rho(0)$) for sufficiently small values of $B$, two solutions  (i.e. two initial values $\rho_1(0)<\rho_2 (0)$) for intermediate values of $B$ (where only the smaller $\rho_1 (0)$ corresponds to a stable solution), and no solution for $B>B_{\rm max}$ (where the value of $B_{\rm max}$ depends on the potential). 

In the MF-EoS case, instead, $\bar\rho (R)$ is nonzero, and no additional condition for $p'(R)$ follows. Different initial values $\bar \rho (0)$ will, therefore, lead to different solutions with different neutron star masses. As is usually done in the TOV approach, we shall assume that solutions are stable as long as  increasing $\bar \rho (0)$ lead to increasing neutron star masses. For practical reasons, we shall, nevertheless, plot both stable and unstable branches in most of the figures, because the numerical integration does not distinguish stable from unstable solutions. On the other hand, the baryon number is no longer a free parameter in the MF-EoS (TOV equation) case. Instead, the baryon number must be determined a posteriori from a given solution via
\be
B = 4\pi \int_0^R dr r^2 \sqrt{\bf B} \bar n_{B}
\ee
where the average baryon density $\bar n_B$ is defined in (\ref{av-b-dens}), and $P$ must be replaced by the TOV solution $p(r)$ in Eq. (\ref{av-b-dens}).  
 
\subsection{Results of numerical calculations}
\begin{figure}
\includegraphics[height=11.5cm]{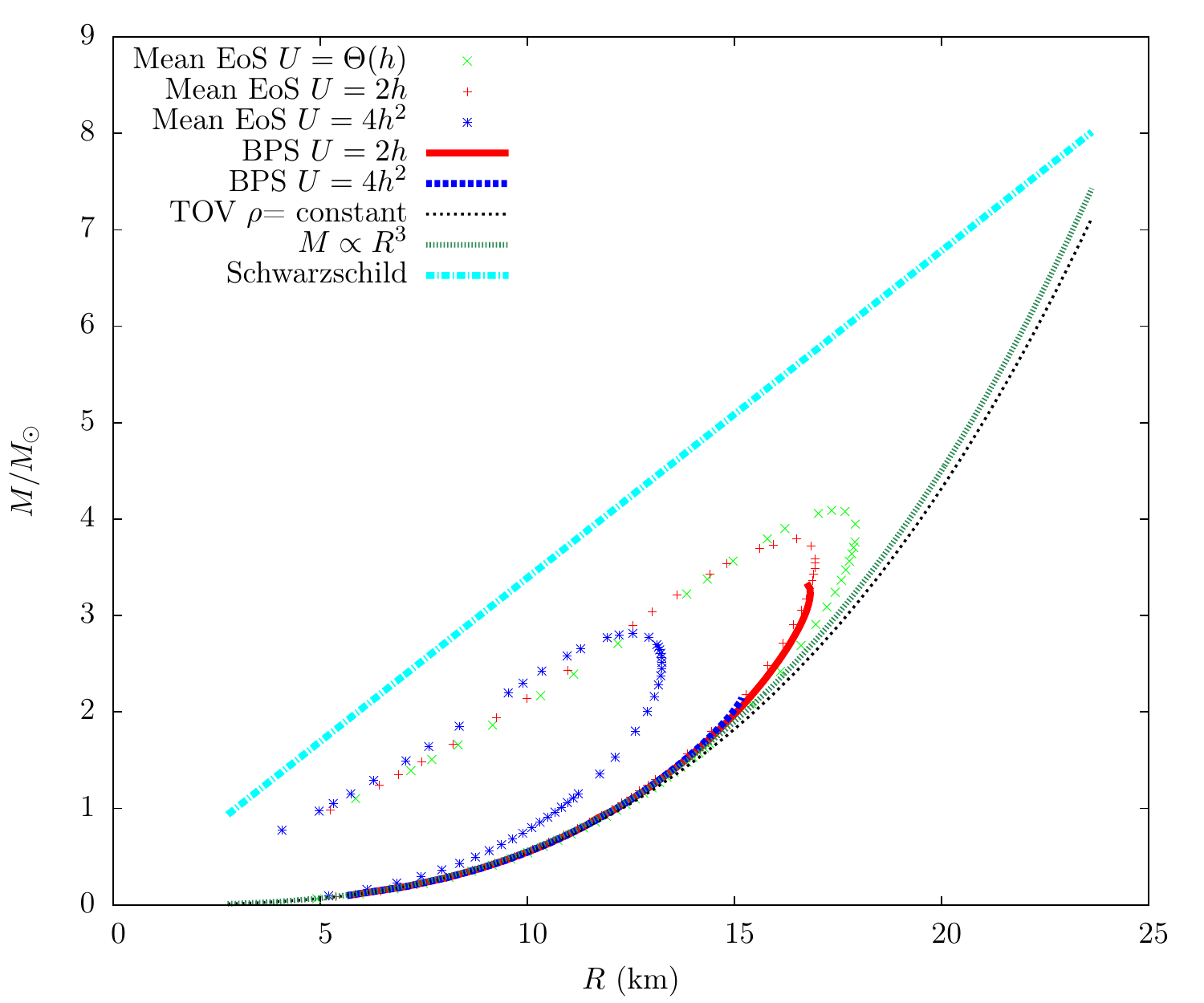}
\caption{(Color online) Neutron star masses (in units of the solar mass) and radii $R$ (in kilometers) for different potentials, both for exact field theory solutions (BPS) and for MF-EoS TOV solutions (Mean EoS). Details are explained in the main text.}
\end{figure}
Numerical solutions are found by a shooting from the center $r=0$. In the MF-EoS TOV case, we impose ${\bf B}(0)=1$ and choose a $\bar \rho (0)= \bar \rho_0$ (which determines $p(0)=p_0$ via the MF-EoS). Then we integrate until we reach a radius value $r=R$ such that $p(R)=0$. The corresponding neutron star mass is then determined from
\be \label{NS-mass}
M=4\pi \int_0^R dr r^2 \bar \rho (r).
\ee
Here, different initial values $\bar \rho_0$ lead to different solutions with different radii and masses. Further, formal solutions exist for arbitrary values of $\bar \rho_0$. We identify stable solutions by the condition that the neutron star mass should grow with growing $\bar \rho_0$, which holds only up to a certain maximum value of $\bar \rho_0$.  

In the full field theory case, we impose ${\bf B}(0)=1$, $h(0)=1$. Then, for each given value of the baryon number $B$, we vary $\rho_0$ until we find a solution which obeys the two conditions $p(R)=0$ and $p'(R)=0$ for some radius $R$. For a fixed value of $B$, we find at most one stable solution. Different neutron star solutions are found for different values of $B$. The neutron star mass is again calculated from Eq. (\ref{NS-mass}) (replacing $\bar \rho (r)$ by $\rho (r)$). 
In this case, it turns out that solutions obeying the two conditions $p(R)=0$ and $p'(R)=0$ only exist up to a certain maximum value of $B$. 

The relations between the resulting neutron star radii $R$ and masses $M$ are shown in Fig. 2. We find that all radii increase with increasing masses (except very close to the maximum masses), both for the exact field theory and for the TOV calculations. This is most likely related to the rather stiff character of the EoS for nuclear matter described by the BPS Skyrme model. For small neutron stars (i.e., for solutions with a small $B$ or $\bar \rho_0$, respectively), we find that the $M(R)$ curve is well approximated by the curve for the EoS $\bar \rho =\mbox{const}$ (by a curve $M\propto R^3$). For the MF-EoS cases, we also show the unstable branches. For these unstable branches of the TOV solutions, we find that in the limit of very large $\bar \rho_0$ they approach a curve $M\propto R$. This is related to the fact that, for all potentials, the MF-EoS approaches $\bar \rho = p$ in the limit of very large $\bar \rho$. 

When comparing exact field theory solutions with MF-EoS TOV solutions for the same potential, we find that the results depend quite significantly on the potential we choose. For the theta function potential, exact field theory calculations and MF-EoS TOV calculations lead to identical results. For the potential ${\cal U}_\pi = 2h$, the two curves are quite similar, the main difference being that the MF-EoS neutron star solution reaches slightly higher neutron star masses. In the case of the ${\cal U}_\pi^2 = 4h^2 $ potential, the difference is more pronounced. The MF-EoS solution not only reaches higher masses, but is also significantly more compact, i.e., the "compactness parameter" $2GM/R$ is significantly bigger than in the full field theory calculation. This difference is probably related to the fact that the potential $4h^2$ is quite peaked about the antivacuum $h=1$ and approaches the vacuum $h=0$ quite fast, i.e., its shape strongly deviates from the constant theta function potential. We plot the compactness of solutions (both full BPS and MF-EoS) for different potentials in Fig. 3.

In contrast to global properties (like the $M(R)$ curves), which are not too different between full field theoretic and MF TOV calculations, the results for local quantities (like the energy densities $\rho (r)$ and $\bar \rho (r)$) are completely different. We plot the energy densities in Fig. 4, the pressures in Fig. 5, and the metric functions ${\bf B}(r)$ in Fig. 6.

\begin{figure}
\includegraphics[height=11.5cm]{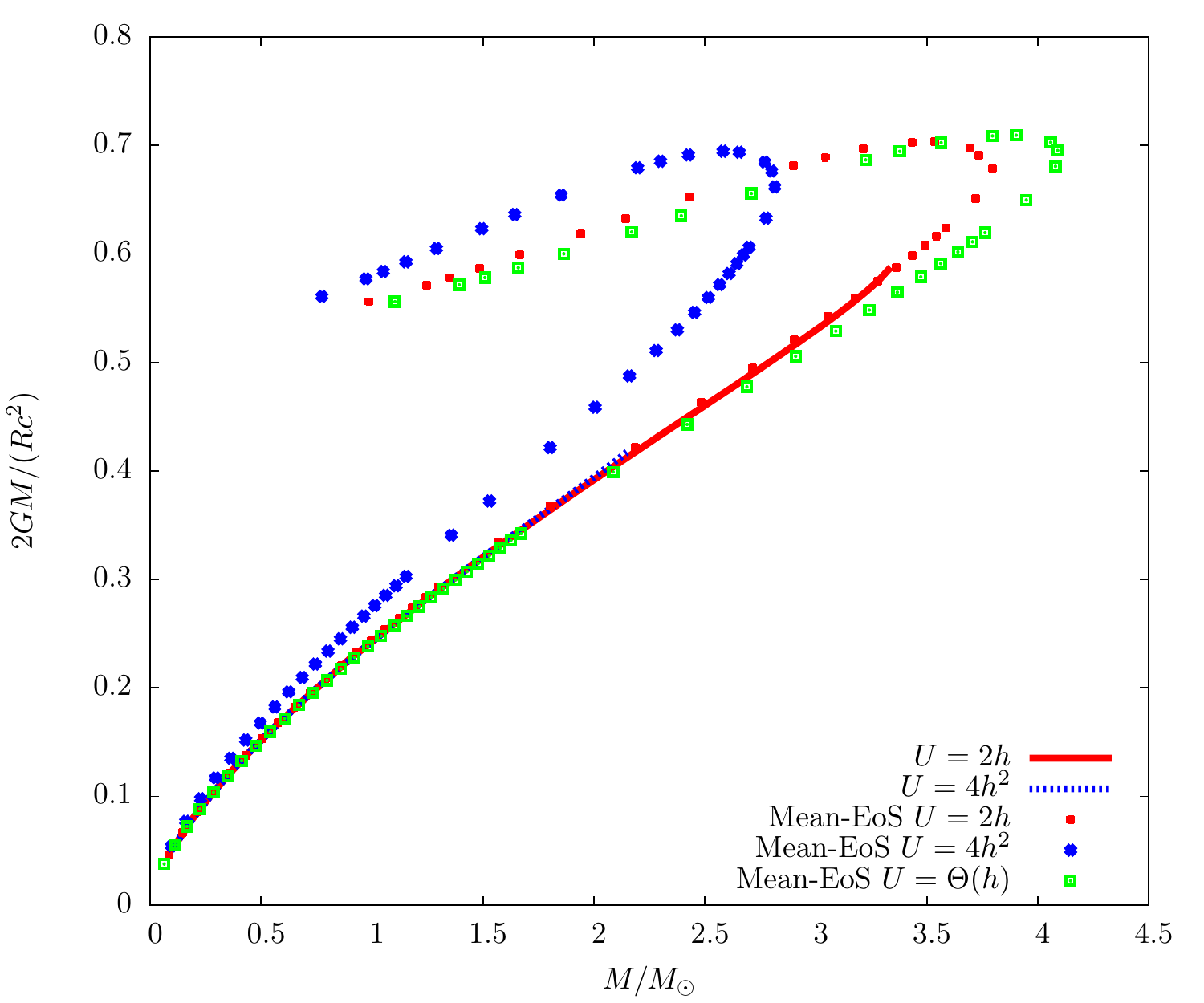}
\caption{(Color online) Compactness of neutron stars as a function of the neutron star masses (in units of the solar mass) for different solutions. For the MF-EoS solutions we also show the unstable branches. It may be seen that in the region of stable solutions all compactness curves are quite similar, with the exception of the MF-EoS solutions for the potential ${\cal U} =4h^2$. }
\end{figure}

\begin{figure}
\includegraphics[height=11.5cm]{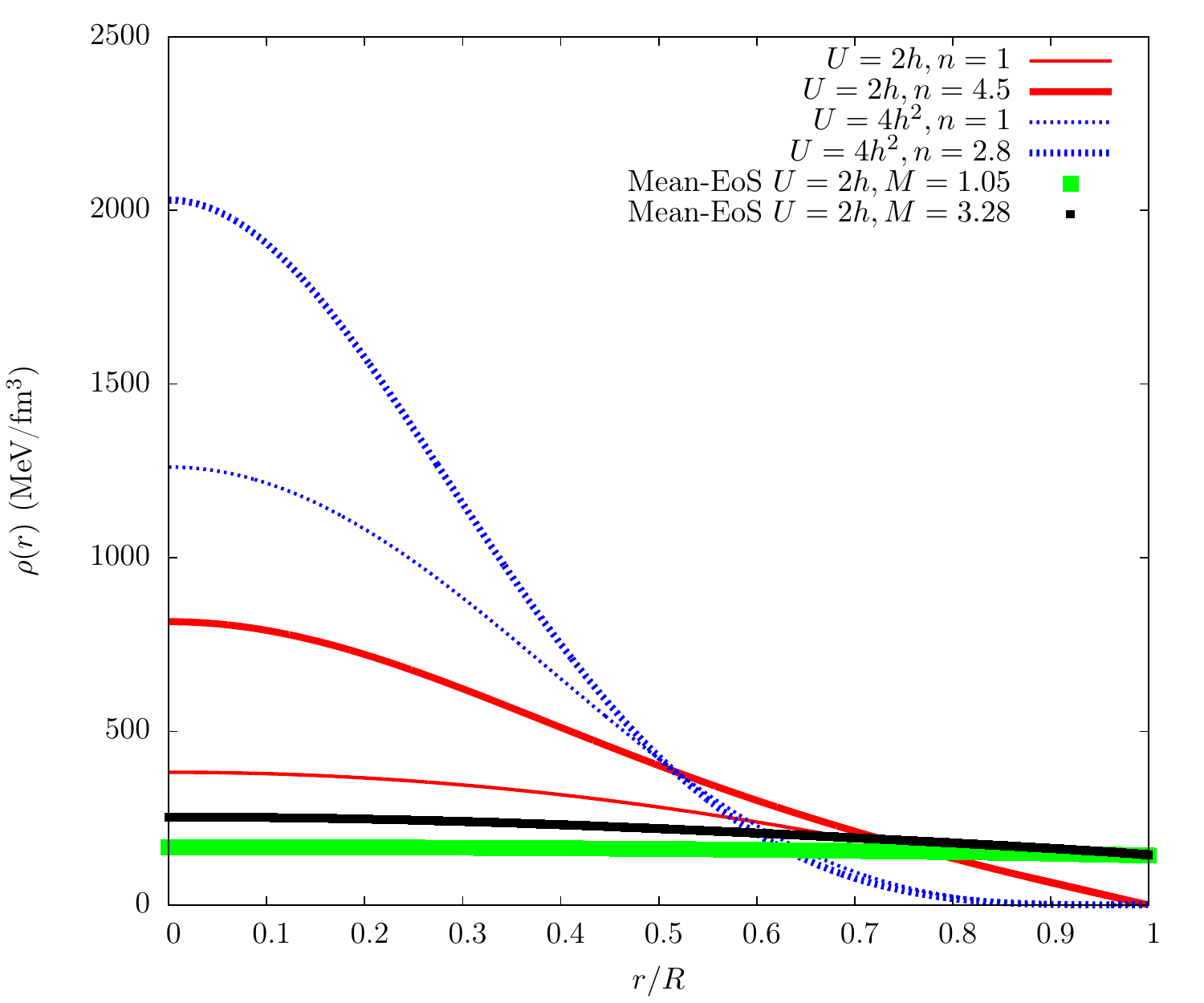}
\caption{(Color online) Energy densities (in MeV fm$^{-3}$) as functions of the radius in units of the neutron star radius $R$. Full BPS energy densities are peaked about the center $r=0$ (especially for heavier neutron stars) and approach zero at the neutron star surface, $\rho (R)=0$. Here, $n$ is the baryon number in solar units, $n=B/B_\odot$. MF-EoS energy densities, on the other hand, vary only slowly and approach a nonzero value (nuclear saturation density) at the neutron star surface. Here, $M$ is the neutron star mass in solar units.}
\end{figure}

\begin{figure}
\includegraphics[height=11.5cm]{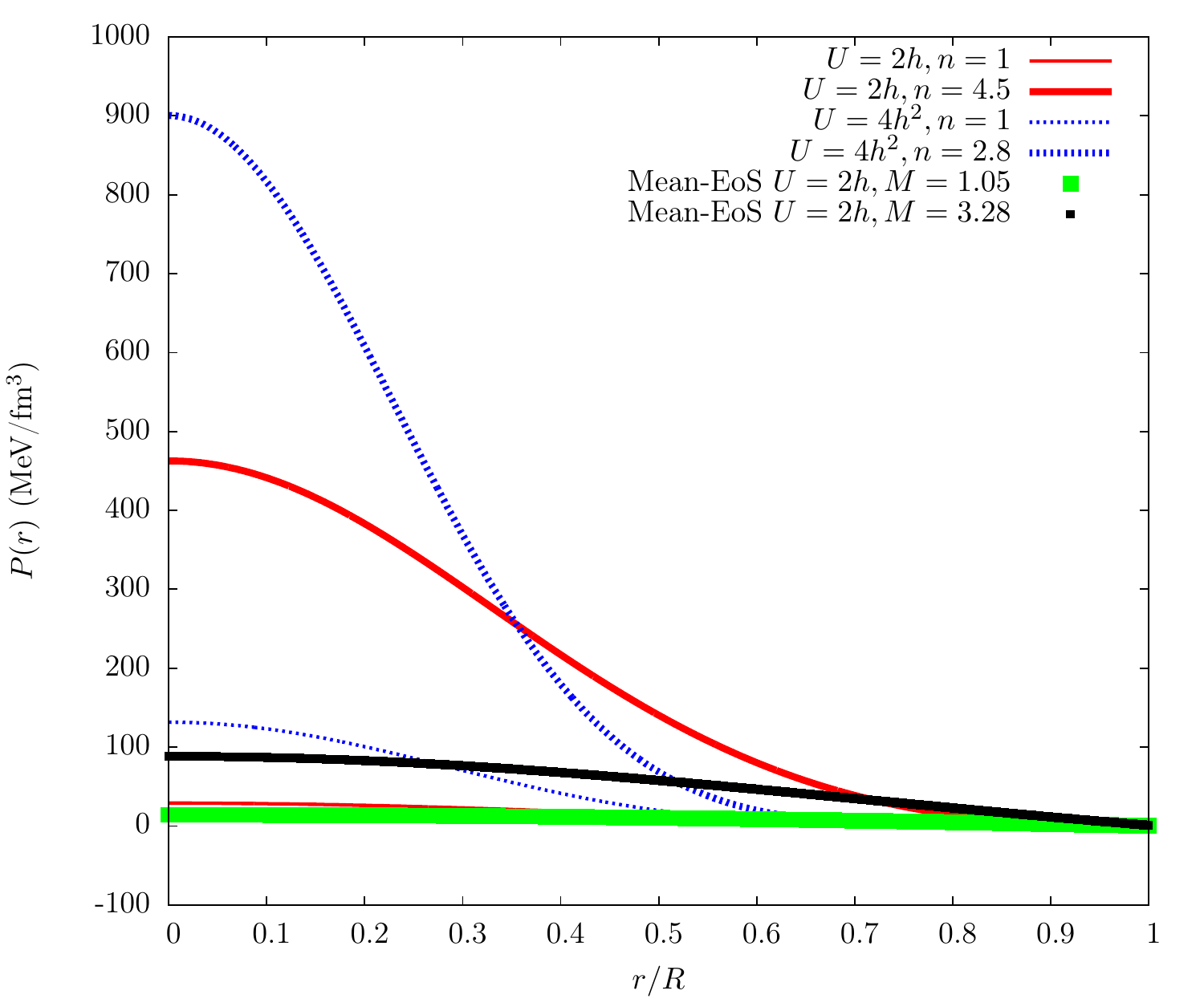}
\caption{(Color online) Pressures (in MeV fm$^{-3}$) as functions of the radius in units of the neutron star radius $R$. Full BPS pressures are peaked about the center $r=0$ (especially for heavier neutron stars). Here, $n$ is the baryon number in solar units, $n=B/B_\odot$. MF-EoS pressures, on the other hand, vary only slowly. Here, $M$ is the neutron star mass in solar units.}
\end{figure}

\begin{figure}
\includegraphics[height=11.5cm]{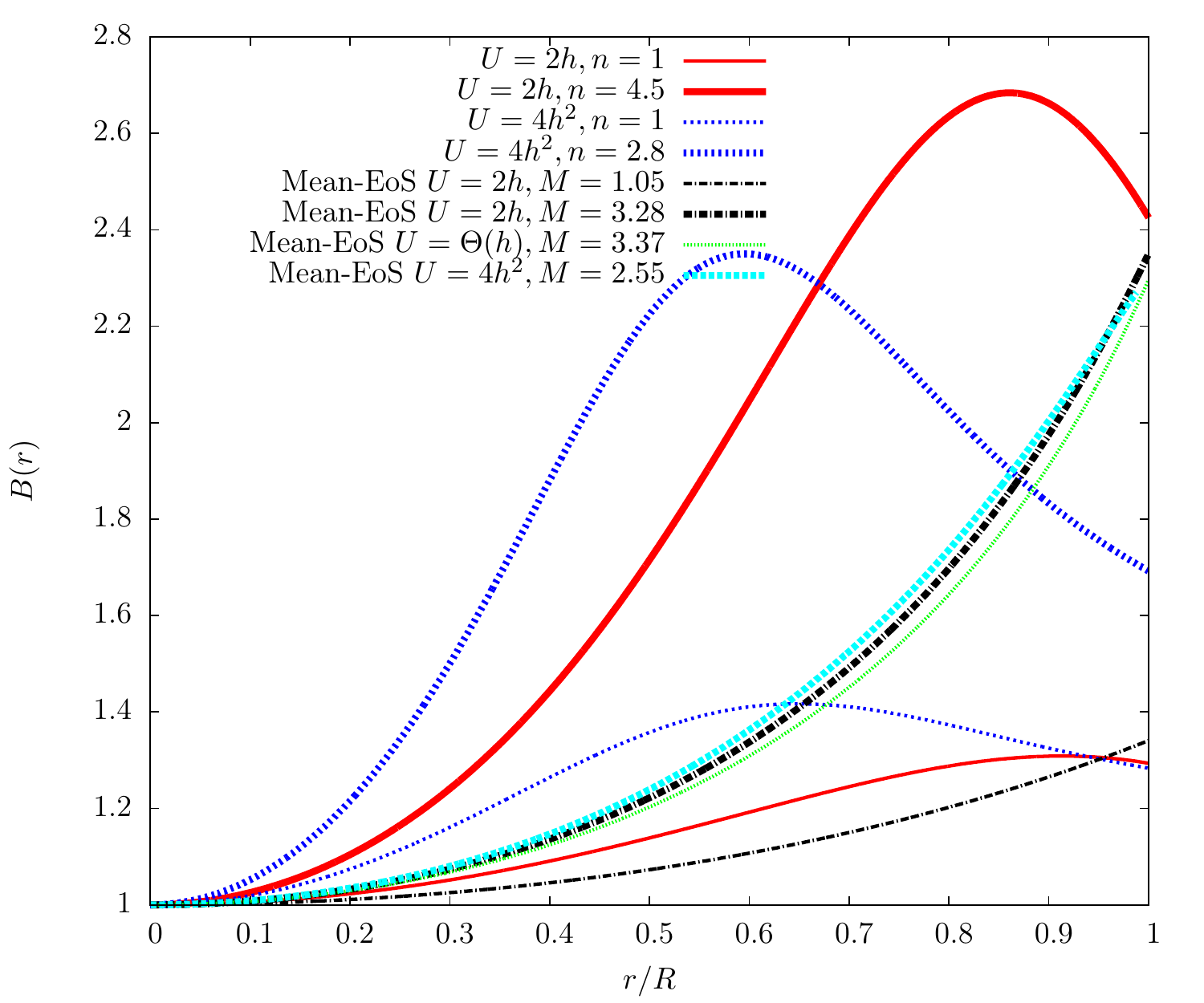}
\caption{(Color online) Metric functions ${\bf B} (r)$ as functions of the radius in units of the neutron star radius $R$. Full BPS solutions ${\bf B}(r)$ have their maxima in the interior of the neutron star, especially for the potential $4h^2$ which has a very peaked energy density. Here, $n$ is the baryon number in solar units, $n=B/B_\odot$. MF-EoS solutions, on the other hand, take their maxima at the neutron star surface $r=R_S$. Here, $M$ is the neutron star mass in solar units.}
\end{figure}

In Fig. 7 and Fig. 8 we show the central values of true and average energy densities and pressures for different neutron star solutions. Again, the difference between true densities (full field theory calculations) and average densities (MF-EoS TOV calculations) is huge. In Fig. 9, we plot the central value of $p$ as a function of the central value of $\rho$.

In Fig. 10, we plot the equations of state (EoS) $p(\rho)$ for different neutron star solutions. All shown EoS are on-shell in the sense that they are reconstructed from the numerical solutions $\rho (r)$ and $p(r)$ by eliminating the independent variable $r$. In the MF-EoS case, the reconstructed on-shell EoS must, of course, coincide with the original MF-EoS. In particular, the on-shell EoS for different solutions corresponding to the same MF-EoS (the same potential) must coincide, and the degree to which they coincide demonstrates the precision of our numerical calculations.
Finally, in Fig. 11 we show the gravitational mass loss of neutron stars for the full BPS model calculations for the two potentials ${\cal U}_\pi$ and
${\cal U}_\pi^2$. More precisely, the figure also includes the (small) mass loss due to the binding energy of infinite nuclear matter. In this paper, we define the solar baryon number $B_\odot$ as solar mass divided by proton mass, 
\be \label{B-sol}
B_\odot = (M_\odot /m_{\rm p}) = ((1.988 \cdot 10^{30} \; {\rm kg}) / (1,673 \cdot 10^{-27} \; {\rm kg})) = 1.188 \cdot 10^{57}
\ee 
(which, strictly speaking, does not exactly coincide with the number of baryon charges in the sun). On the other hand, the neutron star mass approaches $M\sim B E_{B=1}$ in the small mass limit (where gravity can be neglected), where $E_{B=1} = 923,3 \; {\rm MeV}$ is the mass per baryon number of infinite nuclear matter. The ratio $((M/M_\odot)/(B/B_\odot))$ in Fig. 11, therefore, approaches the limiting value $(E_{B=1}/m_{\rm p}) = 0.984$ in the limit of small mass (small $B$).

\begin{figure}
\includegraphics[height=11.5cm]{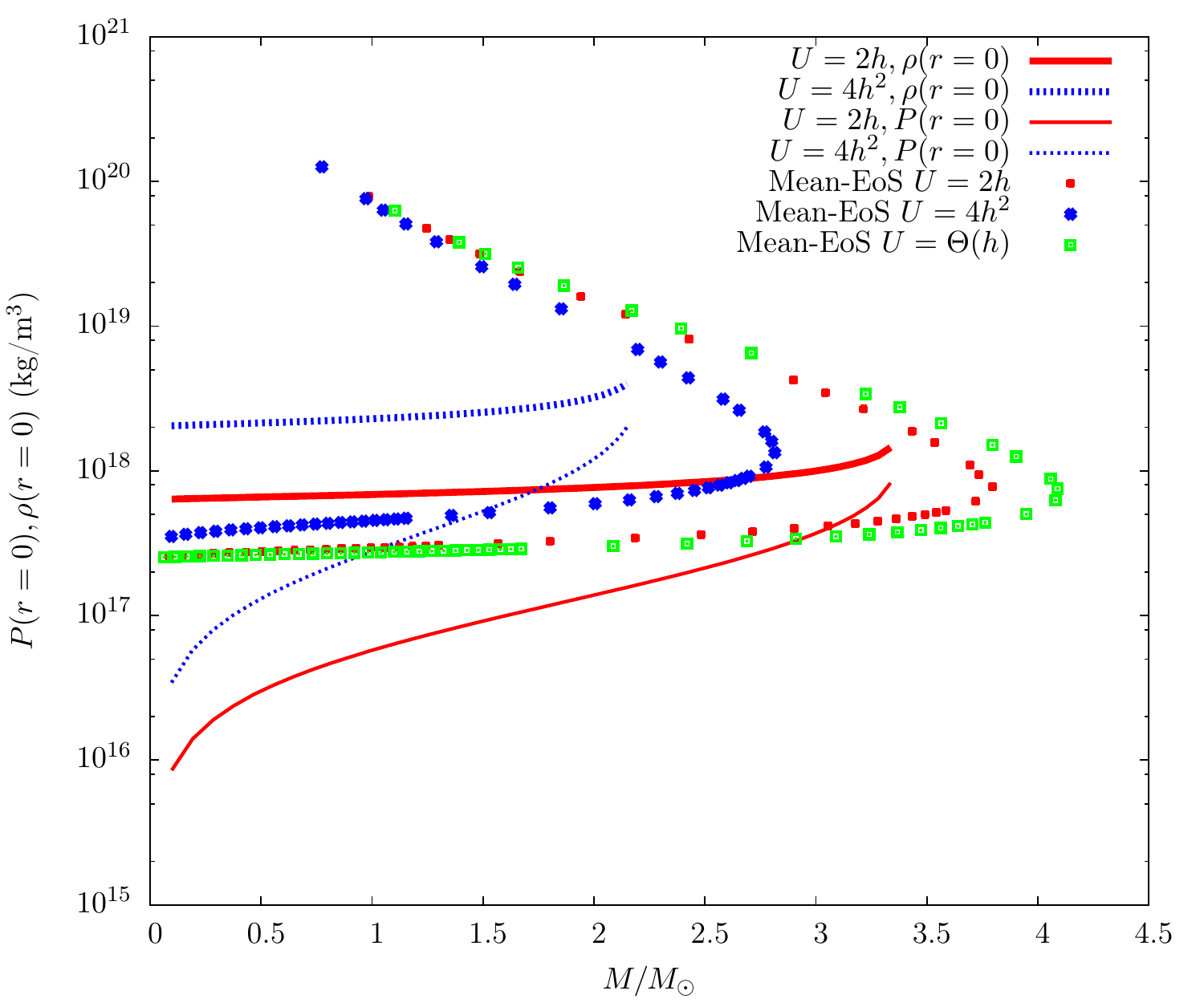}
\caption{(Color online) Central values of the energy density (and of the pressure for the full BPS solutions) for different neutron star solutions, as a function of the neutron star mass. There is a big difference between full BPS solutions, which lead to much larger central values, and MF-EoS TOV solutions. Besides, the central values for the potential $4h^2$ are much larger than for the potential $2h$. In the MF-EoS case, we also show the unstable branches.}
\end{figure}

\begin{figure}
\includegraphics[height=11.5cm]{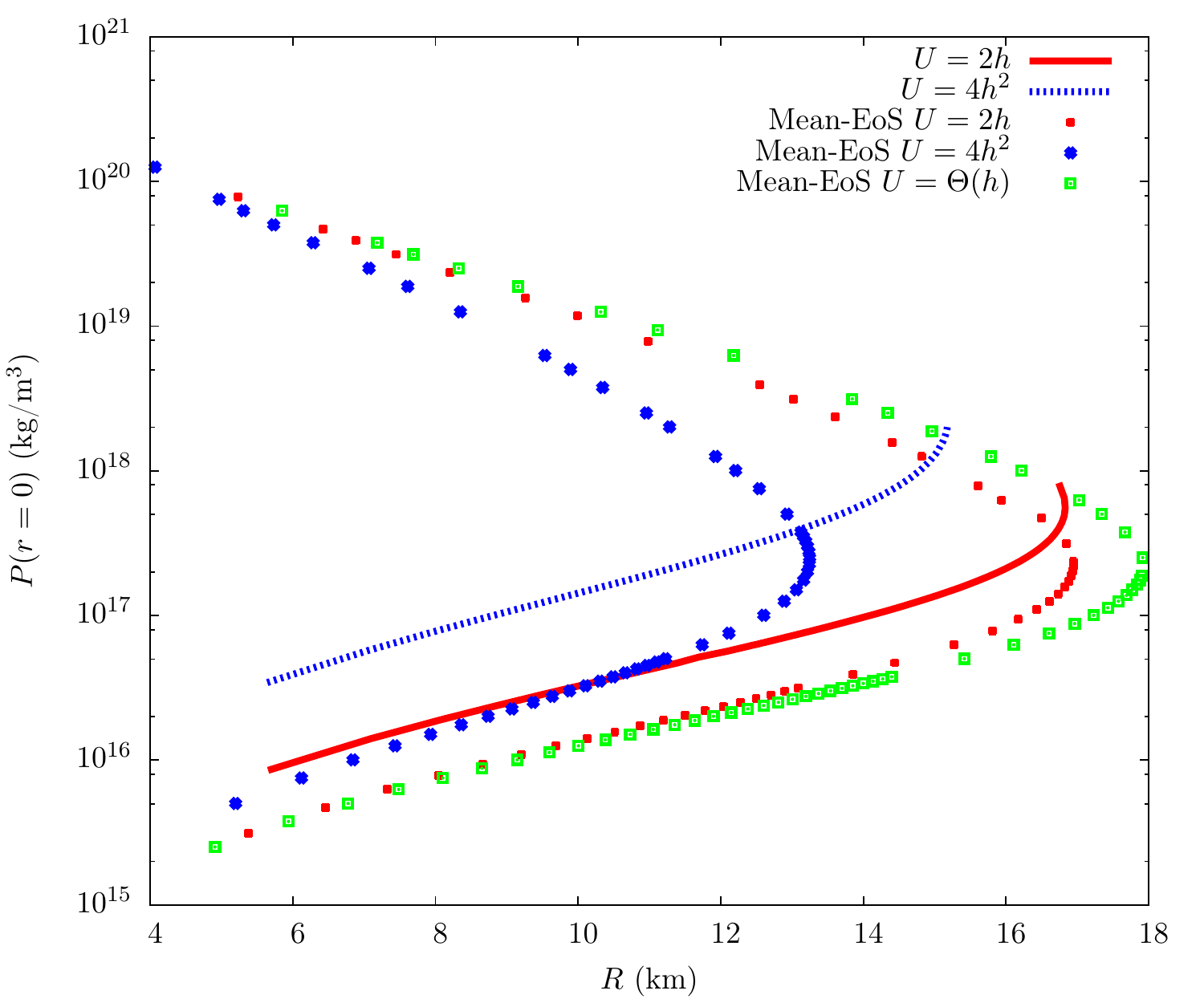}
\caption{(Color online) Central values of the pressure for different neutron star solutions, as a function of the neutron star radius. There is a big difference between full BPS solutions, which lead to much larger central values, and MF-EoS TOV solutions. 
Besides, the central values for the potential $4h^2$ are much larger than for the potential $2h$. In the MF-EoS case, we also show the unstable branches.}
\end{figure}

\begin{figure}
\includegraphics[height=11.5cm]{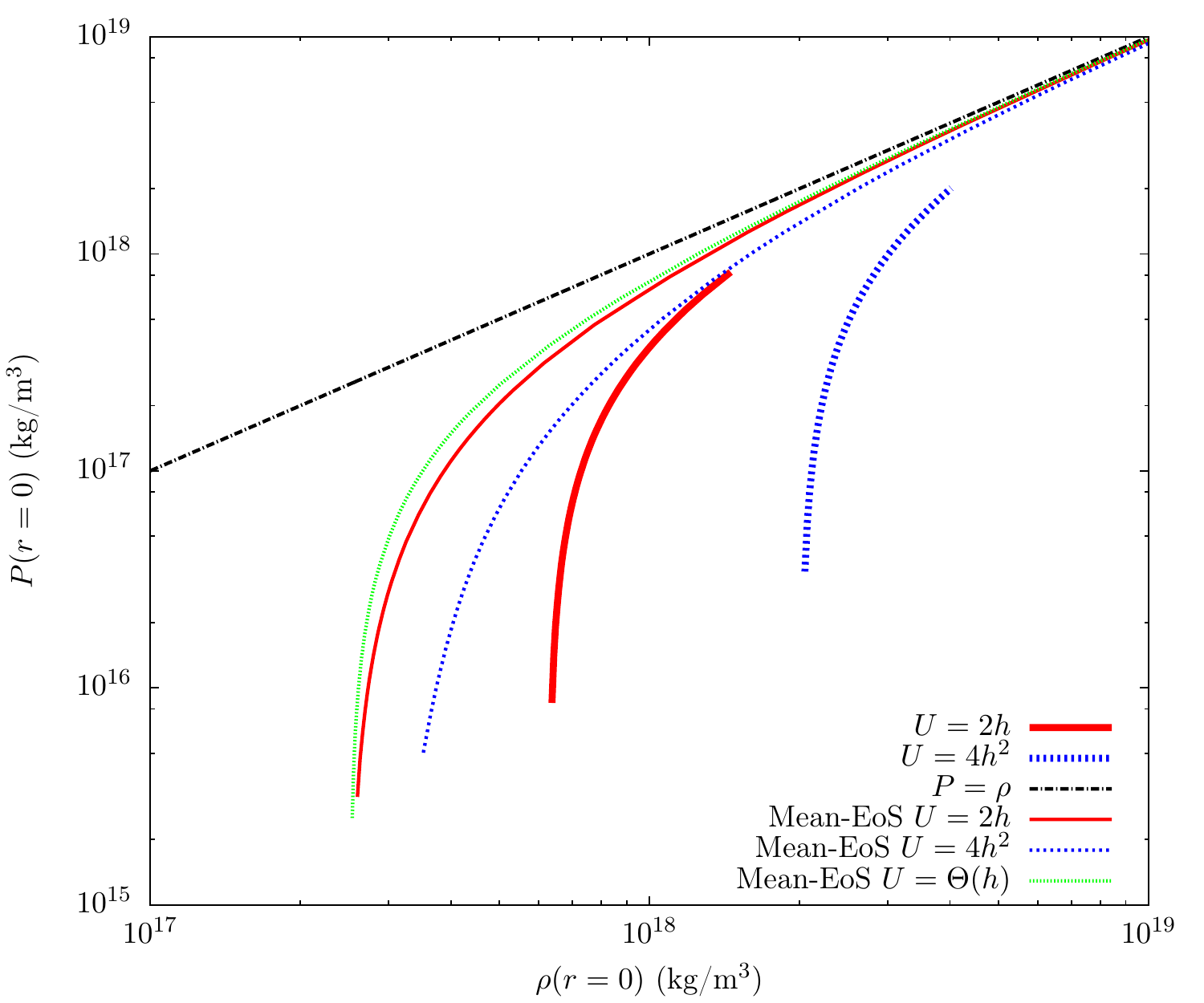}
\caption{(Color online) The central value of the pressure $p(0)$ as a function of the central value of the energy density $\rho (0)$, for different solutions. A given central value of the pressure requires much higher central values of the energy density for full BPS models solutions than for MF-EoS TOV solutions. Observe that, as different solutions for the same potential in the MF-EoS TOV case correspond to the same MF-EoS, the curves for the TOV calculations are, at the same time, the MF-EoS graphs for the corresponding potentials. This is not true for the full BPS model calculations, where different solutions lead to different on-shell EoS even for the same potential, see Fig. 10. }
\end{figure}

\begin{figure}
\includegraphics[height=11.5cm]{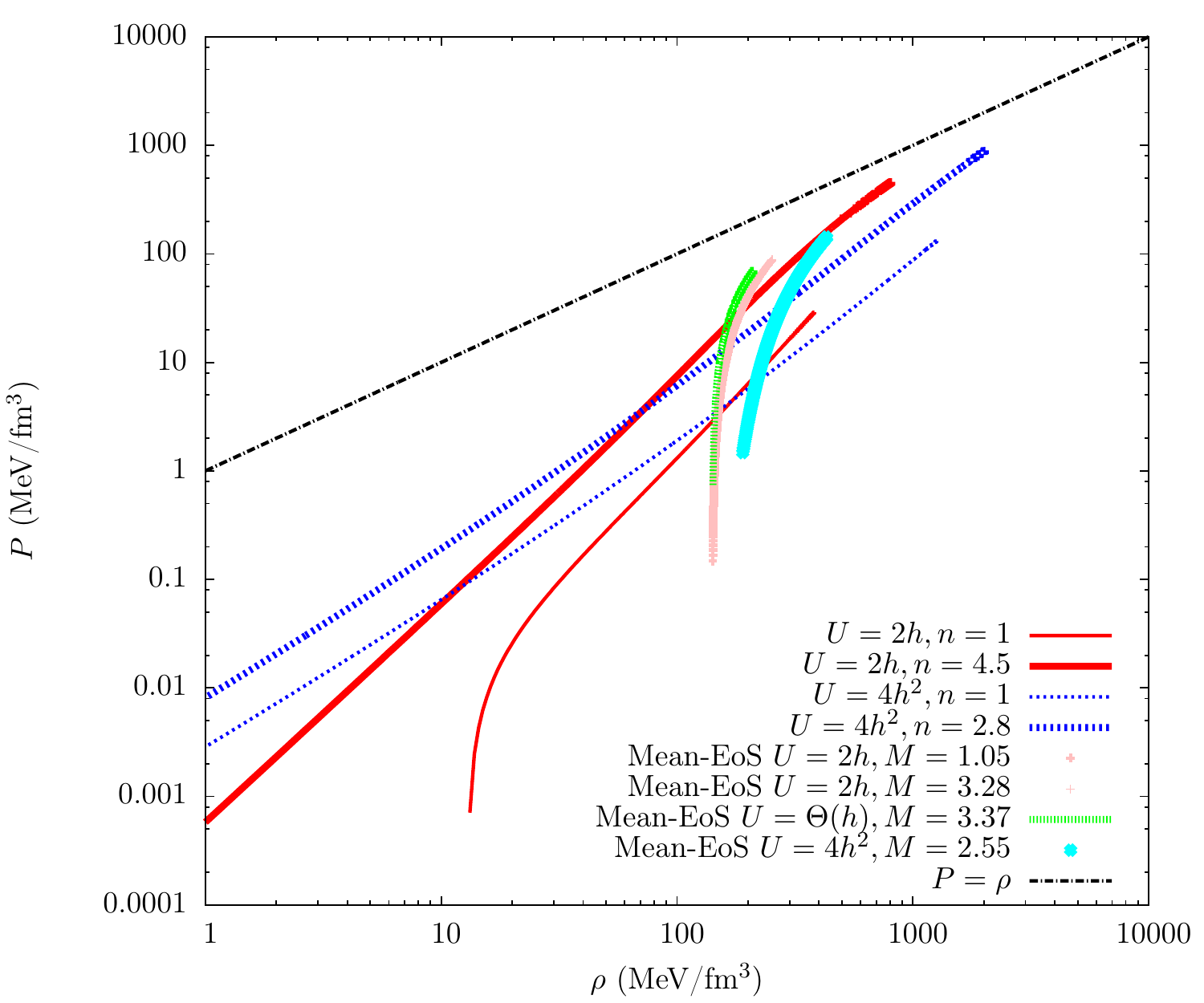}
\caption{(Color online) Equations of state (EoS) for different solutions $\rho (r)$ and $p(r)$. In the case of full BPS model calculations, it is clearly visible that different solutions lead to different on-shell EoS, even for the same potential. Further, the on-shell EoS are approximated with high precision by a homogeneous EoS (polytrope) $p\sim a\rho^b$ where the values of $a$ and $b$ vary even for different solutions for the same potential. The behavior of the MF-EoS, on the other hand, is completely different. In addition, the calculated on-shell EoS for different solutions for the same potential must coincide with each other (and with the original MF-EoS). And indeed, the two different MF-EoS TOV solutions for the potential $2h$ lead to exactly identical numerical EoS in the region of overlapping values of $\rho$ and $p$. Here, $n=(B/B_\odot)$, and $M$ is the neutron star mass in solar units.   }
\end{figure}

\begin{figure}
\includegraphics[height=11.5cm]{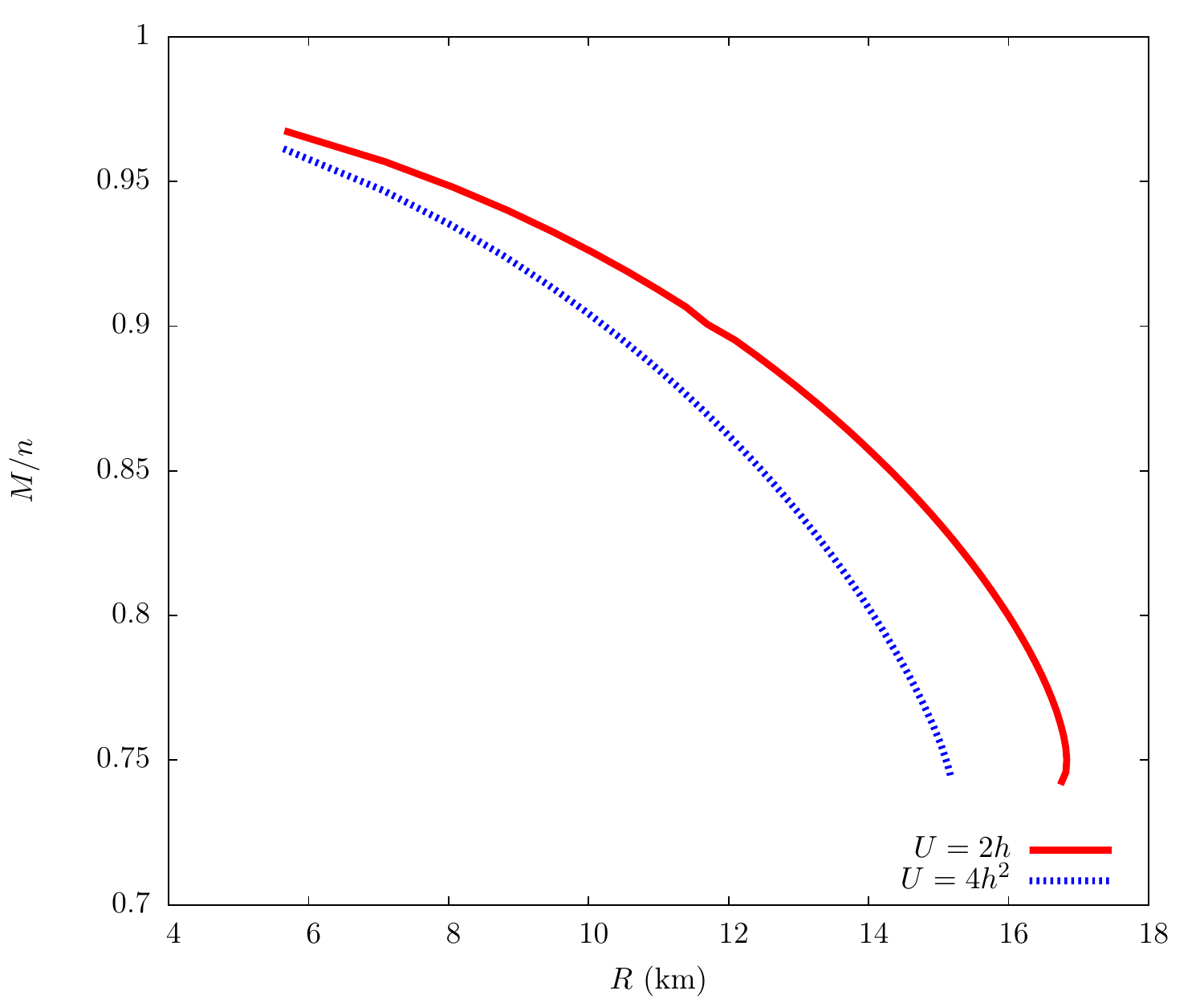}
\caption{(Color online) Gravitational mass loss: we plot the ratio of total mass (with the gravitational mass loss included) to non gravitational (pure neutron matter) mass, $((M/M_\odot)/(B/B_\odot))$, as a function of the neutron star radius, for the full BPS model calculations for the two potentials $2h$ and $4h^2$. The maximum mass loss is about 25\%. Further details are explained in the main text.}
\end{figure}

\section{Discussion of results}
The present work had the two-fold objective to continue, on the one hand, the application of the BPS Skyrme model - a field theory realization of a fluid model of nuclear matter - to the physics of neutron stars, and to investigate, on the other hand, the differences between full field theory calculations and approximate MF calculations of neutron star properties. Concerning the differences between exact and MF calculations, we find that they depend both on the chosen potential and on whether one considers global properties (like the neutron star masses and radii) or local quantities (like energy densities or the metric). There exists one limiting case - the potential $\Theta (h)$ - where MF and exact calculations coincide. For the other two potentials considered, we find that the potential $2h$ leads to rather similar $M(R)$ curves for both exact and MF calculations, where the MF curve reaches, however, slightly larger values for the neutron star mass $M$ (up to about $3.8 \, M_\odot $) than the exact field theory curve (about $3.3 \, M_\odot$), see Fig. 2. For the potential $4h^2$, the difference between the exact and the MF case is much bigger. The $M(R)$ curve for the exact calculation is, indeed, quite similar to the curves for $2h$, but terminates for a smaller mass (about $2.2 \, M_\odot  $; the maximum mass values in \cite{star} are slightly bigger because of the different fit values used there). The curve for the MF calculation for the potential $4h^2$, on the other hand, leads to much more compact neutron stars (smaller radii for the same masses), see Figs. 2 and 3. From our results it is, in fact, easy to understand why the MF result for the potential $4h^2$ is quite different from the MF results for the other two potentials. The MF-EoS for all three potentials rather quickly approach the stiff "hadronic bag type" or maximally compact EoS
\be
\bar \rho = P + B_\infty
\ee
where, however, the numerical values for the "asymptotic bag constant" $B_\infty$ are different for the three potentials. Indeed, using the numerical fit values (\ref{param-pot-theta}), (\ref{param-pot-2}) and (\ref{param-pot-4})  we find 
\bea
\Theta (h): && B_\infty = 2\mu^2 \sim 141 \; {\rm MeV}\, {\rm fm}^{-3} \nonumber \\ 2h: && B_\infty = 2\mu^2 \sim 176 \; {\rm MeV} \, {\rm fm}^{-3} , \nonumber \\
4h^2 : && B_\infty = \frac{5}{2} \mu^2 \sim 353 \; {\rm MeV} \, {\rm fm}^{-3}.
\eea
So $B_\infty$ is much bigger for $4h^2$, which means that for a given value of the pressure $P$ the energy density is much bigger, and the corresponding nuclear matter more compressed, which leads to more compact neutron stars. In the full field theory calculation, it is still true that the matter is more compressed in the center for $4h^2$ but, on the other hand, the potential $4h^2$ approaches the vacuum value $h=0$ faster 
(i.e. takes smaller values near $h=0$)
than the other potentials. This implies that the neutron star (or the soliton in the case without gravity) has a rather large "tail" of low density, which tends to increase its radius.  This shows that the potential $4h^2$ has two properties which have opposite effects on the resulting neutron star radii. On the one hand, this potential is quite peaked about the anti-vacuum $h=1$, which leads to high densities close to the center. On the other hand, the potential $4h^2$ approaches the vacuum $h=0$ fast, which leads to a rather large tail of low density. It would be interesting to consider potentials where these two effects ("peakedness" near $h=1$, and approach to the vacuum near $h=0$) can be varied independently, and to study their influence on neutron star properties. In general, we can say that the differences in global properties (masses, radii) of neutron stars between exact and MF calculations depend on the potential, the most significant difference being that the exact calculations tend to give smaller maximum neutron star masses than the MF calculations. 

For local quantities (like the energy densities, pressures, or metric coefficients as functions of the radial coordinate $r$), the differences between exact and MF calculations are much more pronounced. The true energy densities $\rho (r)$, for instance, take much larger values in the center and approach zero at the neutron star radius $R$, whereas the averaged, MF densities $\bar \rho (r)$ do not vary too much with $r$ (just by a factor of about 2 or 3) and take non-zero values (corresponding to the nuclear saturation density $\rho_s$ of nuclear matter) at the neutron star surface $r=R$. This does {\em not} mean that gravity causes a huge compression for the true energy densities. Instead, the energy densities have a rather similar profile already in the case without gravity (the BPS Skyrme solitons), and, at least for the potentials we considered, the compression induced by gravity is always smaller than by a factor of three in the center, even for the maximum mass cases, see \cite{star}. In other words, the compression at the center in the MF case is not too different from the exact case, although the absolute values of the densities are very different. For the metric functions ${\bf B}(r)$, the most interesting difference is that they take their maximum values inside the neutron star in the exact calculation, but on the surface in the MF calculation, see Fig. 6. These rather big differences for local quantities between exact and MF calculations lead to the plausible conjecture that certain global physical observables with a stronger dependence on the shapes of local quantities will be quite different, too, the most obvious candidate being the moment of inertia relevant for the description of (slow) rotations of neutron stars. Indeed, in the Newtonian case, the moment of inertia is just given by a volume integral of tensorial expressions of the type $x^i x^j \rho (\vec x)$, and the dependence both on the total energy and on the shape of $\rho$ is obvious. In the general relativistic case, however, the moment of inertia is {\em not} just the volume integral of certain moments of the energy density. Instead, its calculation requires to solve the Einstein equations for a more general metric depending on three independent metric functions \cite{mom-inert}, which is beyond the scope of the present article. Still, we expect the Newtonian arguments to be qualitatively valid and, therefore, rather pronounced differences between exact and MF moments of inertia. This issue shall be investigated in detail in a forthcoming publication.

In general, the differences between exact and MF calculations in the BPS Skyrme model will be the stronger the more the potential deviates from the step function potential which provides flat energy and particle densities. Beyond the BPS Skyrme model, this difference should become more important for theories which lead to appreciable inhomogeneities in the energy and particle distributions. In the standard Skyrme model, for instance, one may expect significant variations due to the considerable energy inhomogeneities of the skyrmion crystal \cite{piette3}, \cite{rho}.

Concerning the relevance and usefulness of the BPS Skyrme model (and its near-BPS generalizations) for the description of nuclear matter and neutron stars, we believe that in the present paper we simply added some further strong arguments to an already quite impressive body of evidence.  
Let us briefly repeat some key properties of the model which support this claim. The model
\begin{enumerate}
\item has the energy-momentum tensor of a perfect fluid, so describes a perfect fluid state of nuclear matter.
\item has the SDiff symmetries on physical space among its symmetries, so locally volume-preserving deformations do not cost any energy.
\item has (infinitely many) static solutions (BPS solutions) saturating an energy bound linear in the baryon number (a BPS bound). This allows to easily accommodate the small binding energies of physical nuclei and nuclear matter.
\item incorporates the property of nuclear saturation in the sense that, for arbitrary baryon number $B$, there exist solutions with finite energy and volume (both proportional to $B$) and with zero pressure (the BPS solutions), that is, describing nuclear matter at equilibrium.
\item 
allows for a derivation of its macroscopic, thermodynamical properties directly from the exact, microscopic description (BPS solutions), without the necessity of a thermodynamical limit. 
\end{enumerate}
In the present paper we find, among other new results, that the EoS for a MF averaged energy density $\bar \rho$, for bigger values of $\bar \rho$ and pressure $P$, rather quickly approaches the asymptotic form $\bar\rho = P + B_\infty$. This EoS is called the "maximally compact" EoS in the neutron star literature and, indeed, leads to $M(R)$ curves which are rather similar to the ones we find, although slightly more compact \cite{latt} (the case of the step function potential is, of course, exactly equivalent to the case of the maximally compact EoS, with $B_\infty = B_0 = \rho_s$). For small $P$, on the other hand, our MF-EoS are much softer. Indeed, for potentials with an approach to the vacuum like $\lim_{\xi \to 0} {\cal U} \sim \xi^\alpha$ such that $\alpha \ge 2$ - to which both the potential $2h$ ($\alpha =2$) and the potential $4h^2$ ($\alpha =4$) belong - it follows easily from the results of \cite{term} that the speed of sound at nuclear saturation is zero, $v^2_{P=0} = \lim_{P\to 0} (\partial \bar \rho /\partial P)^{-1} =0$. The general picture we find is that of a maximally stiff (maximally compact) nuclear matter in the core of the neutron star and, more generally, for densities which are sufficiently above nuclear saturation density $\rho_s$. Near nuclear saturation and, therefore, near the neutron star surface, on the other hand, the nuclear matter gets much softer, where the details of this transition between "core" and "mantle" depend on the chosen potential.  Very close to the neutron star surface ("neutron star crust") we believe, in any case, that the BPS Skyrme model is not sufficient, and more terms of the near-BPS model  (\ref{near-BPS})  should be included for a reliable description \cite{star}. 
Some bulk properties of neutron stars (like their maximum masses, or $M(R)$ curves), however, probably do not depend too much on the crust properties, and for these the BPS Skyrme model makes some rather robust predictions, like $M(R)$ curves which are quite similar to the $M(R)$ curves for nuclear matter with the "maximally compact" EoS where, e.g., $(dM/dR)>0$ for almost all neutron stars (except, probably, very close to the maximum mass). This differs from the $M(R)$ curves which result from a large class of nuclear physics EoS (see, e.g., \cite{latt} - \cite{piek}), but is completely compatible with the (still not very precise and not very abundant) observational data. 

More precisely, modifications of the equation of state at and below nuclear saturation (e.g., via a generalization to the near-BPS Skyrme model, or by ad-hoc modifying the EoS at low densities using results from nuclear physics) will have almost no influence on $M$ or $R$ for sufficiently heavy neutron stars. The reason is that these regions of small density are, at the same time, regions of very small pressure, and regions of very small pressure may occupy only the very thin outermost shell of the star, due to the strong gravitational pull of a heavy neutron star. For light neutron stars, on the other hand, the influence of low-density regions on the mass will still be rather small, but their influence on the radius may be more appreciable, leading to larger radii. As of today, however, there is no firm observational evidence for the existence of neutron stars with masses significantly below one solar mass, so the discussion about the "correct" $M(R)$ curve for light neutron stars might well be purely academic.

Another issue where our results may be of some relevance is the so-called TOV inversion. Indeed, for a given EoS for barotropic nuclear matter,  
$p=p(\rho)$, and for a given initial value $\rho (r=0)=\rho_c$ within a certain allowed range, a unique stable neutron star solution with its mass $M$ and radius $R$ follows from the TOV equations. Therefore, by varying $\rho_c$ over its allowed range, the whole $M(R)$ curve may be constructed for a given EoS. The TOV inversion now consists in the inverse operation, i.e., in the reconstruction of a barotropic EoS from a given curve $M(R)$. The formal reconstruction method was developed in \cite{lind}, and combined with statistical methods for an approximate reconstruction of $p(\rho)$ from a finite number of observational data, e.g., in \cite{stein}. Obviously, the TOV inversion hinges on the assumption of barotropic nuclear matter with a barotropic EoS $p=p(\rho)$. The BPS Skyrme model (and certainly also its near-BPS extensions), on the other hand, represents a well-motivated model of non-barotropic nuclear matter. A barotropic EoS may still be found by a mean-field limit (which is well-defined and straight-forward in this model), but the TOV inversion will, nevertheless, be problematic if the $M(R)$ curves of the exact theory differ from the ones for the MF theory. In this case, the EoS reconstructed from an observed $M(R)$ curve (which should correspond to the full field theory) via the TOV inversion will, in general, be different from the EoS resulting from the MF limit. We remark that this is not a specific problem of our model but will be present whenever nuclear matter is described by a field theory which leads to a non-barotropic fluid beyond mean field theory.    

In our model, there exists a formal limit (the limit of very large density) where the difference between exact and MF EoS disappears. Indeed, the exact model has the non-barotropic EoS (off-shell) $p=\rho -2\mu^2 {\cal U}$ where (on-shell) ${\cal U}={\cal U}(r)$ depends on $r$, whereas the MF model EoS soon approaches $p = \bar \rho -B_\infty$, $B_\infty = $ const. Obviously, the difference between the two becomes immaterial in the limit of large $\rho$, or $(\rho -p)/\rho <<1$. This limit is, however, never attained for stable neutron star solutions, for which it is always true that $(\rho_c -p_c)/\rho_c >0.5$ even for the central values.   

\section{Comparison with other results}

In our investigation, we presented an ample variety of both qualitative and quantitative results about neutron stars described by the BPS Skyrme model.
Still, precise quantitative predictions of neutron star properties within our model should be considered preliminary at the present moment. Firstly, as said, we think that a reliable description of nuclear matter requires the inclusion of further terms of the full near-BPS Skyrme model, although the contribution of these further terms to bulk quantities (like neutron star masses or radii) is probably small. Secondly, we fitted the two parameters of our model to the mass per baryon number of infinite nuclear matter, $E_{B=1} = 923.3 \; {\rm MeV}$, and to the nuclear saturation density $n_0 = 0.153 \; {\rm fm}^{-3}$. Here, the first value  $E_{B=1} = 923.3 \; {\rm MeV}$ is quite precise, and almost all nuclear models lead to values $923 \le E_{B=1}/{\rm MeV} \le 926$ (see, e.g., Table 1 in \cite{weber}). The nuclear saturation density, on the other hand, is quite model dependent, where most models lead to a range of values $0.145 \le n_0 \cdot {\rm fm}^3 \le 0.175$ (see, again, Table 1 in \cite{weber}). Thirdly, at the moment, we do not know the "true" potential and so have the freedom to choose different potentials. 

It is, nevertheless, interesting to compare our results with results of neutron star studies within other approaches. In particular, there exist some rather generic results which are either independent of a specific EoS or depend on it only weakly, which makes a comparison all the more relevant.

\subsection{Comparison with generic results}
\subsubsection{The Rhoades-Ruffini bound}
The Rhoades-Ruffini bound is a bound for the maximal neutron star mass originally derived by Rhoades and Ruffini in \cite{RR} which is based on the following observation (a transparent discussion may be found, e.g., in \cite{glend}). Let us restrict to "reasonable" EoS, i.e., EoS satisfying the constraints $0\le dp/d\rho \le 1$ (the pressure increases with the density, and the speed of sound is bound by the speed of light), and let us assume that one reference point $(\rho_f ,p_f)$ of the EoS is known. The result of Rhoades and Ruffini then says that among all EoS the contribution to the neutron star mass for $\rho \ge \rho_f$ is maximal for the maximally compact EoS
\bea
\rho &=& \frac{1}{2}\left[ \rho_f - p_f + (\rho_f + p_f)\left( \frac{n}{n_f} \right)^2 \right] ,  \nonumber \\
p &=& \rho - (\rho_f - p_f) \, , \hspace*{1cm} n \ge n_f
\eea
(here $n$ is the baryon density and $n_f$ its value at the reference point $(\rho_f ,p_f)$).  In a next step one has to choose values $\rho_f$ and $p_f$ for the reference point and an EoS for $\rho <\rho_f$ which smoothly joins the reference point, where both the values for $\rho_f$ and $p_f$ and the low-density EoS should follow from the properties of nuclear matter. Then a numerical value for the maximal mass may be calculated by numerically integrating the TOV equations. In the original paper \cite{RR},  $\rho_f = 4.6 \cdot 10^{17}\, \mbox{kg}\, \mbox{m}^{-3}$ together with the EoS for free degenerate neutrons for $\rho < \rho_f$ was chosen, leading to the bound $M_{\rm max}\le 3.2 M_\odot$. In \cite{glend} a bound is reported for the same value of $\rho_f$ but for an improved EoS for $\rho < \rho_f$, leading to  $M_{\rm max}\le 3.14 M_\odot$. In \cite{glend} also the extreme case $\rho_f = \rho_s = 2,51 \cdot 10^{17} \, \mbox{kg}\, \mbox{m}^{-3}$ is considered, leading to the bound $M_{\rm max}\le 4.3 M_\odot$. 
Here, $\rho_s$ is the nuclear saturation density
\be
\rho_s = 923\, {\rm MeV} \, \cdot \, 0.153 \, {\rm fm}^{-3} =  141\, {\rm MeV}  \, {\rm fm}^{-3} = 
2,51 \cdot 10^{17}\, \mbox{kg}\, \mbox{m}^{-3} .
\ee
If, instead, the maximally compact EoS is used all the way down to $p=0$, i.e., $\rho_f = \rho_s$, $p_f =0$ (so that there is no matching to a soft EoS for $\rho < \rho_f$), then the mass bound is $M_{\rm max} \le 4.09 M_\odot$, see \cite{latt}. But this case is exactly equivalent to the BPS Skyrme model for the step function potential, so it is a gratifying consistency check that the maximal masses in the two cases precisely agree, $M_{\rm max} = 4.1 M_\odot$ for ${\cal U} =\Theta (h)$. We display our maximum masses in table 1. We find that the maximum masses are slightly above the Rhoades-Ruffini bound of $3.2 M_\odot$ for the potential $2h$, but below the bound for $4h^2$. 

\begin{table} \label{tab1}
\begin{center}
\begin{tabular}{|c|c|c|c|c|c|}
\hline
 Potential & \; $\Theta (h) $\;  & $2h$, exact & $2h$, MF & $4h^2$, exact & $4h^2$, MF \\
\hline 
$M_{\rm max}/M_\odot$ & 4.1 & 3.34  & 3.79 & 2.15 &2.82 \\
\hline 
\end{tabular}   
\caption{Maximal neutron star masses for different potentials, both for the exact and the MF (mean field) calculations.} 
\end{center} \label{table1}
\end{table}

\subsubsection{The compactness limit}

The compactness of a neutron star with mass $M$ and radius $R$ is defined as
\be
\beta = \frac{2GM}{R}
\ee
(remember that we use units where the speed of light $c=1$). Obviously, the radius of a neutron star must always be bigger than its Schwarzschild radius $R_S =2GM$, which implies the bound $\beta <1$. Just using relativity and the TOV equations, a tighter bound $\beta <(8/9)$ may be proved \cite{buch}. With additional assumptions, even tighter bounds may be proved. Using the same assumptions on the EoS as Rhoades and Ruffini in their mass bound ($0\le dp/d\rho \le 1$, and a smooth matching to calculable nuclear physics EoS near nuclear saturation), Glendenning derived the improved bound \cite{glend1}
\be \label{glend-bound}
\beta \le \frac{1}{1.47} = 0.68 \equiv \beta_{\rm G}.
\ee
We plot the compacntess values for the maximum mass solutions of our model in table 2.
 \begin{table} 
\begin{center}
\begin{tabular}{|c|c|c|c|c|c|}
\hline
 Potential &\; $\Theta (h) $\; & $2h$, exact & $2h$, MF & $4h^2$, exact & $4h^2$, MF \\
\hline 
$\beta$ for $M_{\rm max}$ & 0.7 & 0.58  & 0.68 & 0.42 &0.66 \\
\hline 
\end{tabular}   
\caption{Compactenss of neutron stars of maximal mass for different potentials, both for the exact and the MF (mean field) calculations.} 
\end{center} 
\end{table}
We find that the step function potential does not satisfy the bound. This is not surprising, because the step function potential leads to the maximally compact EoS for all densities and so cannot be matched to a softer EoS near nuclear saturation. All other potentials are compatible with the Glendenning bound for the compactness parameter. It can be seen that the exact solutions lead to significantly lower values for the compactness parameter than the MF solutions. 

\subsubsection{The energy density limit}
In the TOV approach for a given EoS, a particular solution (together with its resulting neutron star mass $M$ and radius $R$) is determined by the energy density at the center $\rho_c \equiv \rho (r=0)$. For dimensional reasons, the three parameters are related by $\rho_c = \tilde\gamma MR^{-3}$ where, however, even for the same EoS different solutions may lead to different values for the dimensionless constant $\tilde\gamma$. We want to focus, however, on the maximal mass solution, where for a given EoS there exists just one physically acceptable solution leading to one unique relation 
\be
\rho_c = \gamma M R^{-3} .
\ee
Here, the important point is that, for a given EoS, the maximal mass solution provides the maximum value for $\rho_c$ (remember that formal solutions for even higher $\rho_c$ and lower masses are unstable and, therefore, do not correspond to neutron stars). If this expression is now combined with the compactness limit of the previous section (which may be rewritten as a lower bound for the radius, $R\ge (2GM/\beta_{\rm G})$), an upper bound for the central density $\rho_c$ in terms or the maximum mass $M$ follows,
\be
\rho_c \le \left( \frac{\beta_{\rm G}}{2G}\right)^3 \frac{\gamma}{M^2} .
\ee   
For a specific upper bound (a specific value for $\gamma$), one would still have to find the maximum mass solution for a particular EoS. Lattimer and Prakash \cite{latt1} considered, instead, the possibility to derive an universal, EoS-independent bound by using the exact solution of the TOV system known as Tolman VII solution \cite{Tol}. The Tolman VII solution has the energy density
\be
\rho = \rho_c \left( 1-\frac{r^2}{R^2}\right)
\ee
and has the following interesting properties. i) From the above expression, exact solutions may be determined for the metric functions and the pressure, using the TOV equations. ii) Using the scaling symmetry of the TOV equations, $r\to \lambda r$, $m \to \lambda m$, $\rho \to \lambda^{-2} \rho$, $p\to \lambda^{-2}p$, Tolman VII solutions with different values for the parameters $\rho_c$, $R$ and $M$ may be produced from a given solution. Here, $m=m(r)$ is the mass function related to the metric function via ${\bf B}=(1-(2Gm/r))^{-1}$ such that the first TOV equation simplifies to $m' = 4\pi r^2 \rho$. Further, $M=m(R)$.    iii) As a consequence, there exist Tolman VII solutions with arbitrary values for the mass $M$ and radius $R$. This does not contradict the mass bounds derived above, because different Tolman VII solutions do not correspond to the same EoS. iv) There exists, however, a bound $\beta_{\rm TVII}$ on the compactness parameter $\beta$ such that for 
 $\beta \ge \beta_{\rm TVII}$ physically sensible Tolman VII solutions no longer exist. Concretely, $\beta_{\rm TVII}= 0.771 >\beta_{\rm G}$ ($\beta_{\rm G}$ is defined in (\ref{glend-bound})) such that Tolman VII solutions are still valid for $\beta = \beta_{\rm G}$. Consequently, they allow for a rather high compactness and so support high values of $\rho_c$. It is, therefore, quite natural to use the Tolman VII solution for $\beta = \beta_{\rm G}$ as a phenomenological upper limit for $\rho_c$. One easily calculates
 \be
 \rho_c = \frac{15}{8\pi} \frac{M}{R^3} \quad \Rightarrow \quad \gamma = \frac{15}{8\pi}
 \ee
 such that the energy density bound becomes \cite{latt1}
 \be \label{rho-c-bound}
 \rho_c \le \left( \frac{\beta_{\rm G}}{2G}\right)^3 \frac{15}{8\pi  } \frac{1}{M^2} \equiv \rho_{c,{\rm TVII}} \simeq 1.45 \cdot 10^{19} \left( \frac{M_\odot}{M} \right)^2 {\rm kg}\, 
 {\rm m}^{-3} .
 \ee
Lattimer and Prakash checked this bound for a large number of neutron star EoS and found that all values for $M_{\rm max}$ together with their corresponding $\rho_c$ values saturate the bound, and that some values are, in fact, quite close to the bound, i.e., (\ref{rho-c-bound}) is a rather tight bound. 

In table 3 we show the values of $\rho_c$ for the maximum masses for our model for different potentials, together with the values of the bound (\ref{rho-c-bound}). 
\begin{table} 
\begin{center}
\begin{tabular}{|c|c|c|c|c|c|}
\hline
 Potential & \; $\Theta (h) \; $ & $2h$, exact & $2h$, MF & $4h^2$, exact & $4h^2$, MF \\
\hline 
$\rho_c/(10^{18} \, {\rm kg}\, {\rm m}^{-3})$  & 0.753 & 1.45  & 0.774 & 4.01 &1.34 \\
\hline 
$ \rho_{c,{\rm TVII}}/(10^{18} \, {\rm kg}\, {\rm m}^{-3})$ & 0.863 & 1.30  & 1.01 & 3.14 &1.82 \\
\hline
\end{tabular}   
\caption{Central energy densities of neutron stars of maximal mass, together with the Lattimer-Prakash bounds \ref{rho-c-bound}. } 
\end{center} 
\end{table}
We find that, both for the $2h$ and for the $4h^2$ potential, in the case of the exact calculation the central density is somewhat above the Lattimer-Prakash bound. We believe that these potentials behave very reasonable near the vacuum $h=(1/2)\sin^2 \xi =0$ (quadratic or quartic approach to the vacuum, respectively), e.g. from the point of view of pion physics. On the other hand, these potentials are quite spiky (non-flat) also near the anti-vacuum $h=1$, which leads to rather non-flat energy and baryon charge densities (and, consequently, to rather high central densities) already in the case without gravity. 
The violation of the bound might, therefore, indicate that more realistic potentials should be flatter near the anti-vacuum, while maintaining the same approaches to the vacuum. In any case, we want to emphasize that the Lattimer-Prakash bound is of a phenomenological rather than absolute nature. 

\subsection{Skyrme related models}

There have been previous attempts to couple the original Skyrme model to gravity and to use the resulting system to model neutron stars \cite{kli}-\cite{gravSk}. In \cite{kli}, \cite{bizon} spherically symmetric skyrmion configurations (hedgehogs) were coupled to gravity, but it was found that - as in the case without gravity - stable solutions do not exist. In \cite{piette1}, \cite{piette2} rational map ansaetze were considered which are known to provide reasonable approximations for not too high $B$ in the case without gravity. The true minimum energy skyrmions at large $B$, however, are known to be skyrmion crystals, and these crystals were used in the analysis of \cite{piette3}, \cite{nelmes}. More precisely, the EoS of the Skyrme crystal was calculated and the resulting energy-momentum tensor then coupled to gravity. As the Skyrme crystal is not a perfect fluid, there may, in fact, exist two pressures (the radial pressure $p_r (r)$ and the tangential pressure $p_t (r)$) even for the spherically symmetric ansatz. 
Numerically, it is found that for sufficiently heavy neutron stars (above about $1.5 M_\odot$), self-gravitation, indeed, leads to an anisotropic deformation of the Skyrme crystal (e.g. to $p_r (r) \ne p_t (r)$). The maximum mass found in \cite{piette3} is about $M_{\rm max} \sim 1.9 M_\odot$. Further, the EoS is very stiff. The radial speed of sound, e.g., is about $v_r \sim 0.57 c$ already in the Skyrme crystal without gravity, and further grows (but always respecting $v_r \le c$) when self-gravitation is taken into account. A more detailed comparison of the results of \cite{piette3} with our BPS Skyrme neutron star results may be found in \cite{star}. 

A different type of skyrmion fluid was developed by K\"albermann in \cite{kalb} and applied to the description of neutron stars in \cite{ouy1}, \cite{ouy2}. In \cite{kalb}, the Skyrme field was coupled to the dilaton (playing the role of the scalar $\sigma$ meson) and to the (vector) $\omega$ meson. Then a dilute fluid approximation was assumed, where only Skyrme fields which are superpositions of non-overlapping $B=1$ skyrmions (nucleons) are considered. The dilute fluid approximation was partly motivated by the non-existence of spherically symmetric higher $B$ skyrmions, and this motivation is to a certain point superseded by the existence of non-spherically symmetric skyrmions (concretely, the skyrmion crystal in the limit of large $B$). The dilute fluid approximation, however, implies that individual $B=1$ skyrmions only interact via mesons, which allows for a mean-field treatment very similar in spirit to the Walecka model, the main difference being that the nucleons are described by (extended) skyrmions instead of the (point-like) fermions of the Walecka model. In particular, a partition function and a related (barotropic) EoS may be derived in the MF limit. The MF theory is then fitted to the properties of nuclear matter near saturation by choosing the right parameter values for the dilaton potential. In \cite{ouy1}, the skyrmion fluid of K\"albermann in the MF limit was coupled to gravity, and the resulting solutions of the TOV equations were compared to neutron star properties. In addition, in \cite{ouy2}, also the isovector-vector $\rho$ meson was coupled to the Skyrme field, allowing for a different in-medium treatment of protons and neutrons (the in-vacuum properties are still the same for both in this model, because the same $B=1$ hedgehog skyrmion solution is used for both of them). The resulting neutron star $M(R)$ curves are, in fact, quite similar to the ones we find in our model, and the maximum masses are about $3M_\odot$, both for the case with and without the inclusion of the $\rho$ meson, see, e.g. Fig. 2 of Ref. \cite{ouy2}.    

Finally, we want to emphasize once more that the $\omega$ meson is, in fact, implicitly present in the BPS Skyrme model. On the one hand, the sextic term 
${\cal L}_6 = -\pi^4 \lambda^2 {\cal B}_\mu{\cal B}^\mu$ is induced by the coupling of the Skyrme model to the $\omega$ meson in the limit of large meson mass. On the other hand, the MF-EoS of the BPS Skyrme model exactly coincides with the EoS of the Walecka model in the large-density limit, where it is precisely the (repulsive force of the) $\omega$ meson which determines the EoS in that limit. This even allows to match the coupling constants of the $\omega$ meson with those of the BPS Skyrme model, as discussed in Section III.A.

\subsection{The double pulsar J0737-3093}

\begin{figure}
\includegraphics[height=11.5cm]{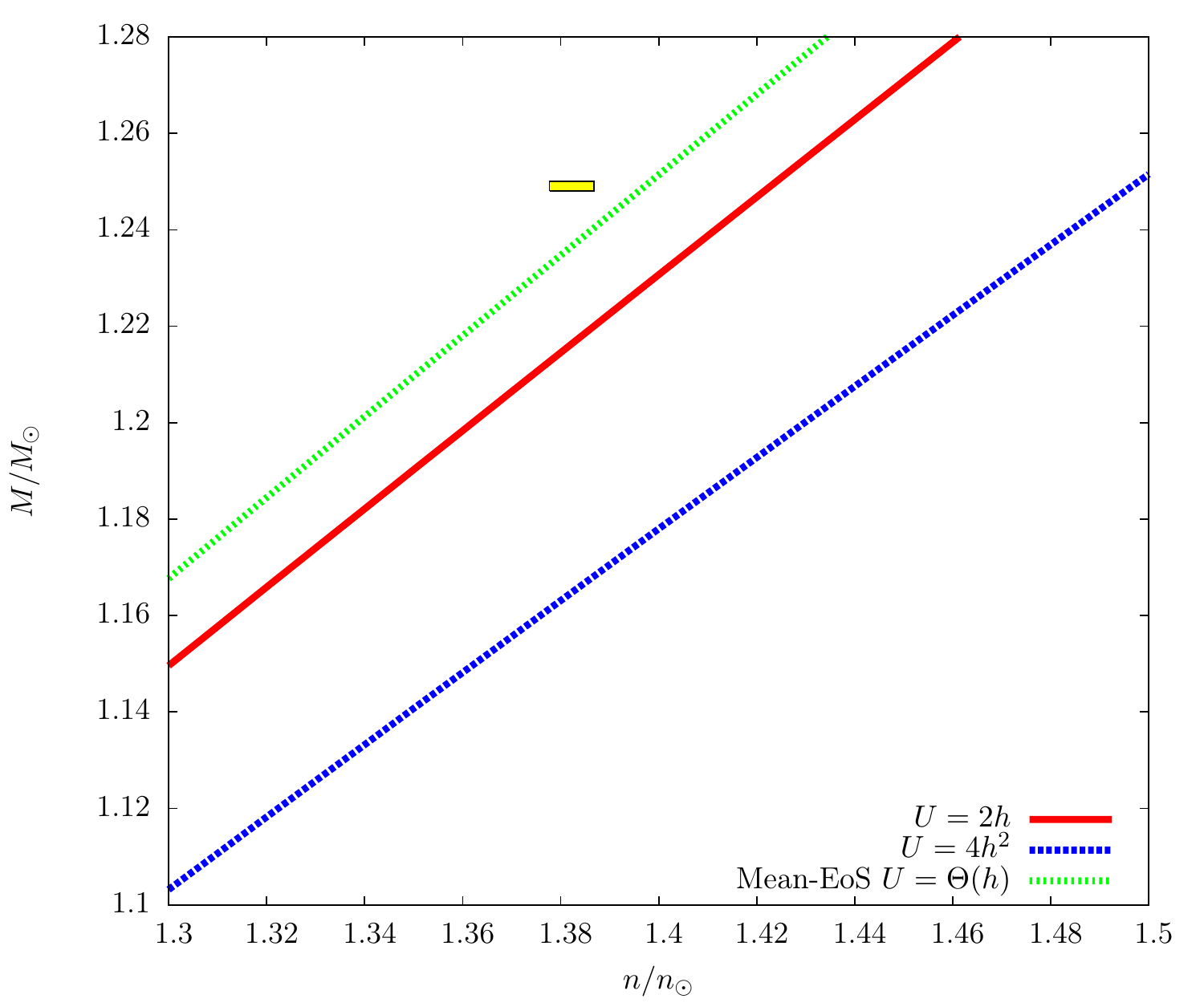}
\caption{(Color online) Comparison of the mass vs. baryon number curves (both in solar units) of the BPS model (exact calculations for $2h$ and $4h^2$, MF-EoS TOV calculation for $\Theta (h)$) with the values determined for the lighter neutron star of the double pulsar J0737-3039 (small yellow rectangle). Details are explained in the main text.}
\end{figure}

Even considering precise predictions preliminary,
it may still be of some interest to consider one particular case where some properties of a specific neutron star are known quite precisely, and to check what our model predicts for this case. Concretely, we refer to the double pulsar J0737-3039 \cite{pods}. The mass of the lighter pulsar ${\rm P}_2$ in this system is determined to be $M_{{\rm P}_2} = 1.249 \pm 0.001 \; M_\odot$ which makes it the lightest firmly established neutron star. If, in addition, it is assumed that this pulsar was formed from an ONeMg white dwarf via an electron-capture supernova - which is plausible given its small mass - then stellar structure calculations allow to determine its baryon number with a rather good precision. Concretely, it is found that the corresponding "baryonic mass" $\bar M_{{\rm P}_2}$ is in the interval $1.366 \le (\bar M_{{\rm P}_2}/M_\odot) \le 1.375$ \cite{pods}. In our model, the baryon number is a much more natural observable than the baryonic mass. Further, in the present paper we defined the solar baryon number $B_\odot$ as solar mass divided by proton mass (\ref{B-sol}). In \cite{pods}, on the other hand, the baryonic mass $\bar M_{{\rm P}_2}$ was determined from the baryon number by assuming a mass per baryon number equal to the atomic mass unit $u=931.5 \; {\rm MeV}$ (which is justified by the most abundant elements in a ONeMg white dwarf). This implies that in the transition from the interval for $  (\bar M_{{\rm P}_2}/M_\odot) $ to the interval for $(B_{{\rm P}_2}/B_\odot) $ we have to multiply by a factor of $(m_{\rm p}/u) = ((938.3 \; {\rm MeV})/(931.5 \; {\rm MeV})) = 1.0073$ leading to the baryon number interval 
$1.376 \le (B_{{\rm P}_2}/B_\odot) \le 1.385$. The resulting small intervals for $M_{{\rm P}_2}$ and $B_{{\rm P}_2}$ may now be compared with the predictions of different models of nuclear matter (i.e., different EoS), see, e.g., \cite{blaschke}. We make this comparison with the $M(B)$ curves of our BPS model in Fig. 12. We find a reasonably good agreement, where especially the potential $\Theta (h)$ gets quite close, which probably means that quite flat potentials are preferred (but remember the caveats at the beginning of this section). 
 
\section{Conclusions and Outlook}

To summarize, we found further evidence for the relevance of the (near-)BPS Skyrme model as a physical model for nuclear matter. The model incorporates in a completely natural fashion many nontrivial qualitative properties of nuclear matter, and, due to its simplicity (at least for the BPS submodel), allows for explicit and at least partly analytical calculations of many properties of nuclear matter, in general, and neutron stars, in particular. Specifically, it makes clear predictions for some neutron star properties, whereas other, more detailed predictions depend on the potential of the model and would, therefore, require a determination of this potential, e.g., from fits to nuclear data. If more precise data on neutron stars will be available in the future, it might even be conceivable to reconstruct the potential (as well as additional terms which are required for a further refinement and completion of the model) from fits to neutron stars, and to use it for further predictions of nuclear properties. We already found some evidence that, while a quadratic or quartic vacuum approach of the potential is expected on theoretical grounds, it should probably be flatter for larger field values (away from the vacuum), leading to flatter energy and baryon density profiles. This issue certainly deserves further investigation.

Further, we found strong indications that MF calculations and exact field theoretic calculations of neutron star properties can lead to different results. If these differences are appreciable, this casts some doubts on the general validity of the TOV inversion, i.e., on the reconstruction of a barotropic EoS $p=p(\rho)$ from an observed $M(R)$ curve. More generally, this implies that if a well-motivated model of nuclear matter does not lead to a satisfactory description of neutron star properties, it might not be the fault of the model but, rather, just indicate the inadequacy of the MF approximation, and the necessity to go beyond MF theory for a reliable description of neutron stars.

On the other hand, the study of rotating neutron stars within the BPS Skyrme model is a rather obvious next step. We believe that the difference between exact and MF calculations will be even more significant in that case. Another important issue is a different treatment of neutrons and protons, which are treated identically in our model, because they correspond to the same classical skyrmion solution. Here, a first possibility is a different in-medium treatment via the introduction of an isospin chemical potential or via a coupling to isovector mesons (like the $\rho$ meson in the dilute skyrmion fluid in \cite{ouy2}). The introduction of a difference between neutrons and protons in Skyrme models already in-vacuum requires the collective coordinate quantization of the isospin collective coordinates. A different treatment of neutrons and protons is certainly required for a realistic description of neutron stars, e.g., to maintain beta equilibrium, or for a consistent coupling to other (nuclear physics) EoS at lower densities. The generalizations of our model necessary to achieve this aim are, however, demanding, since some of the integrability and solvability properties simplifying the analysis of the model will be lost. These generalizations are, therefore, beyond the scope of the present article and will be investigated in forthcoming publications.

\section*{Acknowledgement}
The authors acknowledge financial support from the Ministry of Education, Culture, and Sports, Spain (Grant No. FPA2011-22776), the Xunta de Galicia (Grant No. INCITE09.296.035PR and Conselleria de Educacion), the Spanish Consolider-Ingenio 2010 Programme CPAN (CSD2007-00042), and FEDER. 
CN thanks the Spanish Ministery of
Education, Culture and Sports for financial support (grant FPU AP2010-5772).
AW thanks D. Blaschke and M. Kutschera for comments. AW acknowledge also T. Kl\"ahn for inspiring discussions.

\end{document}